\documentclass[12pt]{article}
\usepackage{amsfonts}
\usepackage{mathrsfs}
\usepackage{amsthm}
\usepackage{multirow}
\usepackage{bbding}
\usepackage{amssymb}
\usepackage{amsmath}
\usepackage{graphicx,color}
\usepackage{enumerate}
\usepackage{comment}
\usepackage{array}
\usepackage{bm}
\usepackage{booktabs}
\newcommand{\bea}{\begin{eqnarray}}
\newcommand{\eea}{\end{eqnarray}}
\newcommand{\Bea}{\begin{eqnarray*}}
\newcommand{\Eea}{\end{eqnarray*}}
\newcommand{\ba}{\begin{array}}
\newcommand{\ea}{\end{array}}
\newcommand{\bt}{\begin{tabular}}
\newcommand{\et}{\end{tabular}}
\newcommand{\btb}{\begin{table}}
\newcommand{\etb}{\end{table}}
\newcommand{\bc}{\begin{center}}
\newcommand{\ec}{\end{center}}
\newcommand{\beq}{\begin{equation}}
\newcommand{\eeq}{\end{equation}}

\newtheorem{defi}{\sc Definition}[section]
\newtheorem{theorem}{\sc Theorem}[section]
\newtheorem{lemma}{\sc Lemma}[section]

\makeatletter

\newcommand{\Rmnum}[1]{\expandafter\@slowromancap\romannumeral #1@}
\makeatother

\begin{document}

\title{Upper expectation parametric regression
\footnote{ The research was
supported by NNSF projects (11171188, 11071145, 11221061 and 11231005) and the 111 project (B12023) of China,  NSF and SRRF projects (ZR2010AZ001
and BS2011SF006) of Shandong Province of China, and a grant from the University Grants Council of Hong Kong. }}
\author{
Lu Lin$^1$, Ping Dong$^1$, Yunquan Song$^1$ and Lixing Zhu$^2$
\\ $^1$Shandong University Qilu Securities Institute for Financial Studies \\Shandong University, Jinan, China\\ $^2$Hong Kong
Baptist University, Hong Kong, China.}
\date{}
\maketitle

\vspace{-4ex}

\begin{abstract} \baselineskip=15pt

Every observation may follow a distribution that is randomly selected in  a class of distributions. It is called the distribution uncertainty.  This is a fact acknowledged in some research fields such as financial risk measure.  Thus, the classical expectation  is not identifiable in general. 
In this paper, a distribution uncertainty is defined, and then an upper expectation regression is proposed, which can describe the relationship between extreme events and relevant covariates under the framework of distribution uncertainty. As there are no classical methods available to estimate the parameters in the upper expectation regression, a two-step  penalized maximum least squares procedure is proposed to estimate the mean function and the upper expectation of the error. The resulting estimators are consistent and asymptotically normal in a certain sense.
Simulation studies and a real data example are conducted to show that the classical least squares estimation does not work and the penalized maximum least squares performs well.

\



{\it Running head:} Upper expectation regression.

\end{abstract}

\baselineskip=21pt

\setcounter{equation}{0}
\section{Introduction}
Suppose that  a sample $\{(X_1, Y_1), \cdots, (X_N, Y_N)\}$ is available in which the observations are independent. 
Consider the parametric regression model:
\begin{eqnarray}\label{(1.1)}
Y_i =g(\beta, X_i) + \varepsilon_i \ \mbox{ for } i=1, \cdots, N,
\end{eqnarray} where $Y_i$'s are the scalar response variables, $X_i=(X^{(1)}_i,\cdots, X^{(p)}_i)^T$'s are  the associated $p$-dimensional covariate with a  probability density $f_X(\cdot)$. The parameters of interest are $\beta$ and the variances of the error. Under the independent identically distributed  case (IID), the  observations  follow a common distribution, the errors $\varepsilon_i$'s then follow a common distribution as well. The estimation and inference about $\beta$ and related parameters have been maturely studied, and 
the estimation consistency and asymptotic normality have been derived in the literature.

However, there are some more complicated scenarios. The IID property may be violated. Heteroscedasticity is one of the scenarios.  A more serious problem is that several factors affect the observations we are obtaining and some of them are often latent, unobservable or
at least unobserved. We call them the latent factors throughout this paper. Very often,
we can not determine exactly how these uncontrolled impacts make the distributions of the observations  different.
Thus, the resulting model structure would be more complicated than the classical heteroscedasticity.  
In this paper, we investigate a more general problem: {\it the distribution of  involved random variable is an element that is randomly selected from a class of  distributions, say $\mathscr F$.} In this sense, all of the elements in $\mathscr F$ can
be seen as possible scenarios in the presence of uncertainty. More specifically, every element $f \in \mathscr F$ can be regarded as a ``conditional" distribution when  latent affecting factor $t\in \cal T$ is given, where $\cal T$ is a set of latent factors. This is called the distribution uncertainty that has been  acknowledged in some scientific fields. We will give a formal definition of distribution uncertainty in the next section. Another relevant methodology is Bayesian statistics, under which the parameter in the distribution is also regarded as a random variable following a prior distribution. However, observations are all with the same parameter value, that is, when the parameter is given, observations are IID. Whereas in distribution uncertainty case, every observation is related to a value of the parameter randomly selected from $\cal T$, the IID property is usually impossible.  A well-known example in mathematical finance and risk measure was raised by a University of Chicago economist Frank Knight (1921).  He distinguished between economic risk and uncertainty through that example. The Knightian uncertainty named after him has then become a well-known notion. Other examples are as follows.
People have a limited ability to determine their own subjective probabilities and might find that they can only provide an interval in which any element can be regarded as a possible probability measure;
as an interval is compatible with a range of opinions, the analysis ought to be more convincing to a range of different people.
Some other relevant researches include  Nutz and Soner (2012) who discussed risk measures under
volatility uncertainty, and the references therein. In probability theory field, Soner et al (2011b)  studied quasi-sure stochastic analysis and Peng (2006) discussed nonlinear expectations and nonlinear Markov chains.

Thus, in this setting, each individual expectation $E_f(Y|X)=g(\beta, X)+E_f(\varepsilon)$ is difficult to estimate from the sample $\{(X_1, Y_1), \cdots, (X_N, Y_N)\}$ because we do not know  which the distribution $f \in \mathscr F$ every observation comes from, even this class of distributions only has finite elements. Under the distribution uncertainty, people often concern the upper expectation of $\mathbb{E}(Y|X)$ that will be defined in (\ref{(1.2)}). This upper expectation can  describe some realistic situations.
For instance, if $Y$ is a measure of the risk of a financial product, the upper expectation regression can describe the relationship between the maximum risk and relevant factors in the sense of averaging. 
A further discussion on the practical application of this model is included  in Section~5 via a real data analysis.  We in this paper  investigate  estimation problems where the distributions of  $\varepsilon_i$'s belong to $\mathscr F$. At the population level, we consider the following upper expectation regression. Assume that $\varepsilon$ follows a distribution $f$ randomly selected from  $\mathscr F$ in a certain sense that will be specified in Section 2.
When $\varepsilon$ is independent of $X$, we have
\begin{eqnarray}\label{(1.2)}
\mathbb{E}[Y|X] =g(\beta, X) +\overline\mu,
\end{eqnarray} where $\overline\mu$ is the upper expectation of $\varepsilon$ defined as
\begin{eqnarray}\label{(1.3)}\overline\mu=\mathbb{E}[\varepsilon]= \sup_{f\in\mathscr F}E_f[\varepsilon],\end{eqnarray} $E_f[\cdot]$ is the classical expectation with a distribution $f$. This upper expectation model is educed from the model (\ref{(1.1)}). 

It is worth pointing out that the notion  of upper expectation is in effect not new
in different settings. It may be at least traced back to, if not earlier, Huber (1981, Chapter 10) in which the relevant topics are related to robust statistics. A non-IID scenario is caused by data contamination and thus every observation still has a fixed distribution, not randomly selected from a class.   
Thus, the scenario is different from the one we are investigating. In this paper, the primary target is  consistently estimating the parameters $\beta$ and $\overline\mu$ by the observations from the model (\ref{(1.1)}) under the distribution uncertainty.  
The following problems  have to be solved in any estimation procedure:
\begin{itemize} \item[{\it 1)}] Upper expectation estimability: The definition of upper expectation implies
the basic feature of nonlinearity, or more precisely, the sub-additivity:$$\mathbb{E}[U+V]\leq \mathbb{E}[U]+\mathbb{E}[V]$$ for any random variables $U$ and $V$. Consequently, the  sample mean when a sample is available cannot be guaranteed to converge to a fixed value such as the classical expectation in IID cases. For example, if the regression function $g(\beta, X)\equiv 0$ in the model (\ref{(1.1)}), then we want to consistently estimate the upper expectation $\overline\mu=\mathbb{E}[Y]$ of $Y$. By Law of Large Numbers (LLN) under sublinear expectation (Marinacci 2005, Peng 2008 and 2009), the sample mean $\overline Y$ of $Y_1, \cdots, Y_n$ would only satisfy that with large probability,
$$\underline\mu\leq \overline Y \leq \overline\mu,$$ where $\underline\mu=\inf\limits_{f\in\mathscr F}E_f[Y]$ and $\overline\mu=\sup\limits_{f\in\mathscr F}E_f[Y]$ are respectively the lower and upper expectation. It presents an obvious evidence that even under very simple models, existing methods have difficulty  to consistently estimate the upper expectation $\overline\mu$. Thus, the estimation consistency is  a very challenging issue under distribution uncertainty.
\item[{\it 2)}] { Data availability:  From problem 1) above, we need to explore the conditions and then select the data that can be used to estimate the parameters of interest.
This raises the issue of data availability. First, intuitively, for any $i$ if $E_f(\varepsilon_i)=\overline\mu$ holds or at least approximately holds in a certain sense, the corresponding observation $(X_i, Y_i)$ could then be used for estimation purpose.
Thus, we have to, under certain conditions, identify those observations. Second,
 a more embedded issue in point estimation is about data availability for different parameters of interest. For instance,  the observations that can be used for  estimating the mean function $g(\beta, \cdot)$ may not be feasible for   estimating $\overline \mu $.

 }\end{itemize}

These problems have not yet been  explored in the literature. The essential difficulties involved in these issues are all rooted in  distribution uncertainty.
The classical statistical methodologies such as the least squires and the maximum likelihood are no longer applicable and thus new method is highly demanded.

     As a useful tool to describe  distribution uncertainty, nonlinear expectation has been developed in the research field of probability theory. 
     A relevant reference is Peng (1997) who introduced   $g$-expectation (small $g$)  via backward stochastic differential equations. As its extension, $G$-expectation (big $g$) and its related versions were proposed by Peng (2006) with $G$-normal distribution as its special case. Related results about LLN and Central Limit Theorem (CLT, Peng, 2008 and 2009) were acquired.  Other references include  Denis and Martini (2006), Denis et al. (2011),
Coquet et al. (2002),
Peng (1999, 2004, 2005),
Nutz (2013), Nutz and Handel(2013). These works offer us a useful foundation in upper expectation research.

In this paper, we consider the class $\mathscr F$ contains finite members. A penalized maximum least squares (PMLS) is introduced and then a two-step estimation procedure is suggested. The key feature of this method is that for different parameters $\beta$ and $\overline\mu$ in the model (\ref{(1.2)}), this method can  identify available data for estimation.
 The resulting estimators are consistent and asymptotically normal in a certain sense. Moreover, the PMLS offers a general estimation approach under the nonlinear expectation framework.

The rest parts of the paper are organized in the following way. In Section~2, the definition of distribution uncertainty is given, the upper expectation regression is reexamined and the motivation for an estimation procedure is  discussed. Section~3  contains the  methodology development, the asymptotic properties of the estimators, the tuning parameter selection and a related algorithm. The method is further extended in Section~4 to the case where the upper expectation can be attained by several distributions. As a special case, the estimator for the upper expectation is constructed in Section~4. Simulation studies and a real data example are presented in Section~5.
The proofs of the theorems are given in Appendix.

\setcounter{equation}{0}
\section{ Definition of distribution uncertainty and motivation}

\subsection{ Definition of distribution uncertainty}
For ease of exposition, we mainly consider the linear case; a brief discussion on the extension of the results to the nonlinear model~(\ref{(1.1)}) will be given at the end of Section 4. In the linear case, the regression function reduces to
\begin{eqnarray}\label{(2.1)} g(\beta, X)=\beta^TX\end{eqnarray}  with
$\beta=(\beta_1,\cdots,\beta_p)^T$ being a $p$-dimensional vector of unknown parameters. What we know is that the distribution of $\varepsilon$ belongs to  a class  $\mathscr F$ so that every $\varepsilon_i$  follows a distribution randomly selected from $\mathscr F$.

To recognize this distribution uncertainty, we give the following definition. 
Suppose that $\mathscr F$ is a distribution class with a factor set $\cal T$ such that $\mathscr F =\{f(\cdot, t): t\in {\cal T}\}.$ On the sample space $\Omega_T$ of $\cal T$, there is a probability measure $P(\cdot)$ such that $p(\cdot)$ is the distribution with respect to $P(\cdot)$. The factor variable $T$ defined on $\Omega_T$ follows the distribution $p(\cdot)$.

\begin{defi}\label{def1}
Let $Z=Z(T)$ be a random variable whose distribution satisfies the following property. For any fixed $T=t \in {\cal T}$, the distribution of $Z=Z(t)$ is $f_t(z)=f(z(t), t) \in {\mathscr F}$ and $T$  is a latent variable following the distribution $p(\cdot)$. We call $Z=Z(T)$ the random variable having the distribution uncertainty in the  class $\mathscr F$. Two random variables are called independent identically distributed under the above distribution uncertainty if they are independent and satisfy the above property.
\end{defi}

This definition can be explained as follows. There is a latent (at least not observed), random  factor(s) $T$ that has impact on the distribution of the random variable $Z$. $\{Z(t): t \in {\cal T}\}$ is a stochastic process/random variable sequence. Consider the pair of random variables $(Z(T), T)$. The corresponding joint distribution is $f(Z(t), t)p(t)$, where $f(Z(t), t)$ can be regarded as a conditional distribution of $Z(t)$ when $T=t \in \cal T$ is given. Because the rendomness of $T$, $Z(T)$ is different from the random variable defined in the classical stochastic process. In the case above, the distribution of $Z(T)$ is uncertain within the class $\mathscr F$ because of the randomness of $T$. If $T$ were observable, the problem would reduce to the classical functional data framework  where all  observations $(Z_i(T_i), T_i)$ were functional data. However, under the distribution uncertainty framework, this is not the case, what we can observe is just $Z_i(T_i)$ in which $T_i$ is latent (or not observed). Therefore, any  element $f(\cdot, t)$ within the class $\mathscr F$  could be the distribution of $Z(T)$  in the above random manner. Without notional confusion, we then simply write $Z(T)$ as $Z$ throughout the rest of the paper. Thus, for a random variable function $g(Z)$, the expectation $E_{f_t}[g(Z)]$ is actually the conditional expectation with conditional density $f_t=f(Z(t), t)$ given above.

In the following, we mainly consider  distribution uncertainty of the error term $\varepsilon$ in model~(\ref{(1.1)}). In this case, the error $\varepsilon=\varepsilon (T)$ is of distribution uncertainty as $Z(T)$ we have defined above. By (\ref{(1.2)}) and (\ref{(2.1)}), the upper expectation linear regression is defined as:
\begin{eqnarray}\label{(2.2)}
\mathbb{E}[Y|X]=\beta^T X+\overline\mu,\end{eqnarray}
where $\overline\mu= \sup_{t\in \cal T}E_{f_t}[\varepsilon (t)]$.
That is to say, the upper expectation is  the maximum of conditional expectations over all $f(\cdot, t) \in {\cal F}$. If we use the conditional expectation notation, we may regard the expectation $E_{f_t}[\varepsilon (t)]$ as a conditional expectation $E_{f_t}(\varepsilon(T)|T=t)$ if $f_t$ is regarded as the conditional distribution of $Z(T)$ when $T=t$ is given.
Note that under the distribution uncertainty, the original model (\ref{(1.1)}) has no a constant intercept term and every expectation $E_{f_t}(\varepsilon_t)$ of $\varepsilon_t$ is not assumed to be zero. Distribution uncertainty causes the expectation $E_{f_t}(\varepsilon)=E_{f_t}(\varepsilon(T)|T=t)$ is a function of the random factor of $T$ and is not always zero. Thus, we need to handle the upper expectation $\overline\mu= \sup_{t\in \cal T}E_{f_t}[\varepsilon (t)]$. On the other hand, there is no need to consider a constant  intercept term as it is not identifiable. In model (\ref{(2.2)}), the intercept is absorbed in $\overline\mu$.

To estimate  $\beta$ and $\overline\mu$, we first suggest an estimation procedure at the population level. A natural objective function is  the squared upper expectation loss:
\begin{eqnarray}\label{(2.3)}\mathbb{E}
\left[(Y-\beta^T X-\overline\mu)^2\right].\end{eqnarray}
We first analyze what is the minimizer of this loss over $\beta$. Because   $\varepsilon$ and $X$  are independent and $X$ follows a certain distribution $f_X$, it is easy to see that the true $\beta$ is the minimizer over all $\beta$. For all $\overline\mu$, we check what can be the minimizer.  When $\beta$ is the true value, it is easy to see that the above squared upper expectation loss is equal to
\begin{eqnarray}\label{(2.4)}\mathbb{E}
\left[(\varepsilon-\overline\mu)^2\right].
\end{eqnarray}
Suppose that there is a distribution $f_{t^*} \in {\mathscr F}$ or  equivalently a factor $t^* \in {\cal T}$ such that the above supremum can be attained over all $f_t$. That is to say,
there exists a member $t^*\in\cal T$ such that \begin{eqnarray}\label{(2.5)}\mathbb E\left[(\varepsilon (T)-\overline\mu)^2\right]=E_{f_{t^*}}\left[(\varepsilon (t^*)-\overline\mu)^2\right].\end{eqnarray} This is a commonly used assumption for identification, in spirit the same as that in Peng (2008 and 2009).  
Then, by the projection theory, it is easy to see that the minimizer of the loss over $\overline\mu$ is $E_{f_{t^*}}(\varepsilon)$, rather than $\sup_{t\in \cal T}E_{f_t}[\varepsilon (t)]$.
Therefore, we need a two-step procedure to estimate $\beta$ and $\overline\mu$ consistently. First, use the above criterion to get $\beta$ and $\overline\mu$. As described above, the estimator $\widehat \beta$ of $\beta$ can be consistent, while the one of $\overline\mu$ cannot. After $\widehat \beta$ being obtained, we then re-estimate $\overline\mu$ to get the estimation consistency. In the following, for ease of presentation, let ${\cal T}=\{1, \cdots L\}$ for a positive integer $L$. Under this situation, $T$ follows a distribution $P_T$ with  unknown probability mass $p_t$ for $t\in\{1, \cdots L\}$.

\subsection{Motivation for estimating $\beta$ and $\overline\mu$ }
 Recall that $(X_i,Y_i):i=1,\cdots,N$  are independent observations from the model:
\begin{eqnarray*}Y_i=\beta^T X_i+\varepsilon_i, \ \ i=1,\cdots,N.\end{eqnarray*} Because of  distribution uncertainty, every realization $\varepsilon_i=\varepsilon_i(t_i)$  has a distribution $f_{t_i}\in \mathscr F$ with the latent factor $t_i$ having the distribution $P_T$. For  given $t_i$'s, we have the linear expectations $\mu_i=E_{f_{t_i}}(\varepsilon_i)$ and variances $\sigma^2_i=E_{f_{t_i}}[(\varepsilon_i-\mu_i)^2]$. The expectations and variances are in effect the conditional ones when the latent random factor $T=t_i$ are given. 
We consider the following  treatment to get the initial estimates of $\beta$ and $\overline\mu$.

For any given $\beta$ and $\overline\mu$, let $\left\{G_{(j)}(\beta,\overline\mu)=(Y_{k_j}-\beta^T X_{k_j}-\overline\mu)^2:j=1,\cdots,N\right\}$ be the ordered quantities of $\left\{G_{i}(\beta,\overline\mu)=(Y_{i}-\beta^T X_{i}-\overline\mu)^2:i=1,\cdots,N\right\}$ in descending order:
\begin{eqnarray}\label{(2.6)}G_{(1)}(\beta,\overline\mu)\geq G_{(2)}(\beta,\overline\mu)\geq \cdots \geq G_{(N)}(\beta,\overline\mu).\end{eqnarray}
To construct an empirical version of $\mathbb{E}\left[(Y-\beta^T X-\overline\mu)^2\right]$, instead of using all $G_{i}(\beta,\overline\mu)$'s to get an overall average, only using those larger $G_{(i)}(\beta,\overline\mu)$'s would make it possible to achieve estimation consistency. The intuition is as follows. Note that  $\mathbb{E}\left[(Y-\beta^T X-\overline\mu)^2\right]$ is the upper expectation being achieved at the distribution $f_{t^*}$. 
Although we do not know what $t^*$ is, 
we can understand that the relatively larger quantities should be close to this upper expectation.
More particularly, it can be expected that there should exist a positive number $n<N$ such that
most of $G_{(j)}(\beta,\overline\mu),j=1,\cdots,n$, come from the distribution $f_{t^*}$. For illustration, we consider a simple example:


{\it Example}. Suppose that ${\mathscr F}=\{f, f_{*}\}$, in which the density functions $f\sim U(-1,3)$ and $f_{*} \sim U(0,4)$, two uniform distribution densities. It can be seen that $E_{f_*}(Z^*)=2> E_f(Z)=1.$
Let ${\cal Z}=\{Z_1,\cdots,Z_n\}$ and ${\cal Z}^*=\{Z^*_1,\cdots,Z^*_n\}$ be the samples of $f$ and $f_*$, respectively. Denote the largest order statistic $U_M=\max\{U_1,\cdots,U_n\}$ and the random event $$A_k=``k \mbox{ elements } Z^*_{i_j}\in {\cal Z}^*\mbox{ satisfy } Z^*_{i_j}\leq U_M,j=1\cdots,k".$$ Let $m=[n^{\delta}]$ for a constant number $0<\delta<1$, and $p=P(Z^*_{i}\leq U_M)$, where $[x]$ stands for the integer part of $x$. Then \begin{eqnarray}\label{(2.7)}P\left(\bigcup_{k=m}^nA_k\right)\leq\sum_{k=m}^nC_n^kp^k(1-p)^{n-k}
\sim\frac{(n-m)n^n}{m^m(n-m)^{n-m}}\frac{3^m}{4^n}
\rightarrow 0 \ \ (n\rightarrow\infty),\end{eqnarray} $\Box$

The proof for (\ref{(2.7)}) will be given in Appendix.
This example shows that in the mixing sample $\{Z_1,\cdots,Z_n, Z_1^*,\cdots,Z_n^*\}$, most of the data that have larger values should come from $f_*$ when $n$ is large enough. It gives a clear evidence to ensure that there exists a number $n$ such that most of $G_{(j)}(\beta,\overline\mu),j=1,\cdots,n$, come from the distribution $f_{t^*}$.
Based on the above observation, for constructing an empirical version of $\mathbb{E}\left[(Y-\beta^T X-\overline\mu)^2\right]$, the following
partial sum seems to work:
\begin{eqnarray}\label{(2.8)}\frac 1n\sum_{j=1}^nG_{(j)}(\beta,\overline\mu) \ \mbox{ for some positive integer } n\leq N.\end{eqnarray}
By this intuition, an estimate $(\beta_n^T,\overline\mu_n)$ of $(\beta^T,\overline\mu)$ would be defined as the minimizer of the partial sum:
\begin{eqnarray}\label{(2.9)}(\beta_n^T,\overline\mu_n)=\arg\min_{\beta\in\mathscr B,\overline\mu\in \mathscr U}\frac 1n\sum_{j=1}^nG_{(j)}(\beta,\overline\mu),\end{eqnarray} where $\mathscr B$ and $\mathscr U$ are  respectively the parameter spaces of $\beta$ and $\overline\mu$.

However, the problem described here is rather more complicated. First, in the probability sense, there are $n=p_{t^*}N$ data points that come from the distribution $f_{t^*}$ for some $t^*\in {\cal T}$. As we do not know the value of $p_{t^*}$, the integer $n_*$ is unknown in practice. Thus, we  cannot have prior information on  what elements in the set $\{G_{i}(\beta,\overline\mu):$ $i=1,\cdots,N\}$ can be included in (\ref{(2.8)}).  How to identify such elements is the key for selecting available data such that the estimation consistency can be achieved. Second,  more seriously, the consistency of $\frac 1n\sum_{j=1}^nG_{(j)}(\beta,\overline\mu)$ to $\mathbb{E}\left[(Y-\beta^T X-\overline\mu)^2\right]$ cannot automatically result in the consistency of $\overline\mu_n$ to $\mu$. This is  because $f_{t^*}$ is to make the loss function possible to achieve the maximum, but the linear expectation $E_{f_{t^*}}[\varepsilon]$ may  not  be equal to the upper expectation $\overline \mu$ of $\varepsilon$.  This shows the complexity of the problem:  for estimating different parameters, the corresponding available data sets may be different.  In the next section, we will give the detail of a two-step estimation procedure.

\setcounter{equation}{0}
\section{Methodology and theoretical properties}


\subsection{First-step estimation of $\beta$ and $\overline \mu$}
In the previous section, we have discussed the issue of sample size determination because we cannot use all of the data for estimating $\beta$ and $\overline\mu$. We now suggest a general method for data selection and parameter estimation simultaneously. To this end, we need to assume that  the distribution $f_*:=f_{t^*}$ exists. 

According to the correspondence between the indices $(j)$ and $k_j$ via $G_{(j)}(\beta,\overline\mu)=(Y_{k_j}-\beta^T X_{k_j}-\overline\mu)^2$ for $j= 1,\cdots,n$, we decompose the index set $I_n=\{k_j:j=1,\cdots,n\}$ into two subsets as $U_n=\{u_j:j=1,\cdots,[n/2]\}$ and $L_n=\{l_s:s=n-[n/2]+1,\cdots,n\}$ satisfying $u_j>l_s$. More precisely,
\begin{eqnarray}\label{(3.1)}I_n=U_n\cup L_n, \mbox{ where } U_n\cap L_n=\emptyset, \mbox{ and } u_j>l_s \mbox{ for any } u_j\in U_n, l_s\in L_n.\end{eqnarray}
Denote
\begin{eqnarray*}\Delta_{n}=\frac {1}{[n/2]}\sum\limits_{j\in U_n}
E[(Y_{j}-\beta^T X_{j}-\overline\mu)^2]-\frac {1}{n-[n/2]}\sum\limits_{j\in L_n}E[(Y_{j}-\beta^T X_{j}-\overline\mu)^2].\end{eqnarray*} 
Since the sums in $\Delta_{n}$ are based on the original indices, instead of the ordered quantities $G_{(j)}(\beta,\overline\mu)$, it can be showed that if most of $(Y_{k_j}-\beta^T X_{k_j}-\overline\mu)^2,j=1,\cdots,n$, come from the distribution $f_*$, then $|\Delta_n|$ should be small enough. Moreover, we need the following condition:
\begin{itemize}\item[{\it C0}.] The scatter plots of $(Y_j-\beta^T X_j-\overline\mu)^2,j=1,\cdots,N$, are asymmetric. \end{itemize} Under this condition, we have that $|\Delta_n|\nrightarrow 0$ if
most of $(Y_{k_j}-\beta^T X_{k_j}-\overline\mu)^2,j=1,\cdots,n$, do not come from the distribution $f_*$. Combining the observations above,
we choose a tuning parameter $\tau>0$ and consider a constraint as $|\Delta_{n_\tau}|<\tau$, where $n_\tau$ depends on $\tau$.
If $|\Delta_n|$ is given, the estimator of $(\beta^T,\overline\mu)^T$ can be defined as
\begin{eqnarray}\label{(3.2)}\left(\widehat\beta^T,
\widehat{\overline\mu}\right)^T=\arg\min\limits_{\beta\in\mathscr B,\overline\mu\in\mathscr U,n_\tau\in\mathscr N}\frac {1}{n_\tau}\sum\limits_{j=1}^{n_\tau}G_{(j)}(\beta,\overline\mu) \ \ \mbox{ s.t. } \  |\Delta_{n_\tau}|<\tau,\end{eqnarray}  Because the expectation of $\frac {1}{n_\tau}\sum\limits_{j=1}^{n_\tau}G_{(j)}(\beta,\overline\mu)$ is a decreasing function of $n_\tau$, the  ideal choice of $n_\tau$ is $$n_\tau=\max\left\{n:|\Delta_{n_\tau}|<\tau\right\}.$$ The relation between $\tau$ and $n_\tau$ implies that the optimization problem (\ref{(3.2)}) contains two tuning parameters $\tau$ and $n_\tau$.
By the Lagrange multiplier, the optimization problem (\ref{(3.2)}) can be rewritten as
\begin{eqnarray}\label{(3.3)}\left(\widehat\beta^T,
\widehat{\overline\mu}\right)^T=\arg\min\limits_{\beta\in\mathscr B,\overline\mu\in\mathscr U,n_\lambda\in\mathscr N}\frac {1}{n_\lambda}\sum\limits_{j=1}^{n_\lambda}G_{(j)}(\beta,\overline\mu)+\lambda |\Delta_{n_\lambda}|,\end{eqnarray} where $\lambda$ is a
tuning parameter,
and the related tuning parameter $n_\lambda$ is related to $|\Delta_{n_\tau}|$. Since $\frac {1}{n_\lambda}\sum\limits_{j=1}^{n_\lambda}G_{(j)}(\beta,\overline\mu)$ is a decreasing function of $n_\lambda$, and $|\Delta_{n_\lambda}|$ is not small when the value of $n_\lambda$ exceeds a certain amount,
the above objective function is an approximate convex function of $n_\lambda$ in a certain region. Also it can be directly verified that the above objective function is a convex function of $\beta$ and $\overline \mu$. As a result, the resulting estimator is a unique global
solution of the above optimization problem.

However, $\Delta_{n}$ depends on unknown expectations $E[(Y_{j}-\beta^T X_{j}-\overline\mu)^2]$ for $j\in U_n\cup L_n$. We need a consistent estimator to replace  it.
Let $\widehat\beta_{LS}$ be the ordinary least squares estimator of $\beta$ based on all of the data $(X_j,Y_j), j=1,\cdots,N$. Denote
\begin{eqnarray*}&&\Upsilon_n^1(X,Y)=\frac {1}{[n/2]}\sum_{j\in U_n}(Y_{j}-X^T_{j}\widehat\beta_{LS})^2-\frac {1}{n-[n/2]}\sum_{j\in L_n}(Y_{j}-X^T_{j}\widehat\beta_{LS})^2, \\ &&\Upsilon_n^2(Y,\overline\mu)=\frac {1}{[n/2]}\sum_{j\in U_n}(Y_{j}-\overline\mu)^2-\frac {1}{n-[n/2]}\sum_{j\in L_n}^{n}(Y_{j}-\overline\mu)^2.\end{eqnarray*} We then have the following conclusion.

\begin{lemma}\label{lemma3.1}
{\it Assume that  $\varepsilon_i$'s are independent of $X_i$'s and the variances $\sigma_i^2$ of $\varepsilon_i$  with distribution $f_i$ exist for all $i=1,\cdots, N$, then
$$\Delta_n=\Upsilon_n^2(Y,\overline\mu)-2\overline X^T\beta\Upsilon_n^1(X,Y)
+O_p\left(1/\sqrt{n}\right),$$ where $\overline X=\frac 1n\sum_{i=1}^nX_i$.}\end{lemma}

This lemma leads to that the optimization problem (\ref{(3.3)}) is asymptotically equivalent to
\begin{eqnarray}\label{(3.4)}\nonumber &&\hspace{-6ex}\left(\widehat\beta^T,\widehat{\overline\mu}\right)^T
\\&&\hspace{-6ex}=\arg\min_{\beta\in\mathscr B,\overline\mu\in\mathscr U,n_\lambda\in\mathscr N}\frac {1}{n_\lambda}\sum_{j=1}^{n_\lambda}G_{(j)}(\beta,\overline\mu)+\lambda\left|\Upsilon_{n_\lambda}^2(Y,\overline\mu)-2\overline X^T\beta\Upsilon_{n_\lambda}^1(X,Y)\right|.\end{eqnarray} Here $\Upsilon_n^2(Y,\overline\mu)-2\overline X^T\beta\Upsilon_n^1(X,Y)$ replaces $\Delta_n$.  
For any given $\overline\mu$ and $\beta$, a  choice of $n_\lambda$ is $$n_\tau=\max\left\{n:\left|\Upsilon_n^2(Y,\overline\mu)-2\overline X^T\beta\Upsilon_n^1(X,Y)\right|<\tau\right\}.$$
The above estimation method is called the penalized maximum least squares (PMLS). Under $G$-normal distribution (see Peng 2006), it is a penalized maximum-maximum likelihood. 
 The penalty used here is to control the difference between the second-order moments of the random variables and then to identify the available data set. 

We now investigate the theoretical properties of the estimators of $\beta$ and $\overline \mu$ in (\ref{(3.4)}). Denote $\mathscr G_{n}=\{G_{(1)}(\beta,\overline\mu),\cdots,G_{(n)}(\beta,\overline\mu)\}$ and suppose that there are only $d_n$ elements  $G_{(j_s)}(\beta,\overline\mu),s=1,\cdots, d_n$, in the set $\mathscr G_{n}$ such that $G_{(j_s)}(\beta,\overline\mu),s=1,\cdots, d_n$, do not come from $f_*$. Let $\mathscr G_{n_0}$ be the smallest set of $\mathscr G_{n}$ that contains all the elements $G_{(j)}(\beta,\overline\mu)$ from the distribution $f_*$. Suppose without loss of generality that when $n\geq n_0$, only the last $d_n$ elements $G_{(n-d_n+1)}(\beta,\overline\mu),\cdots, G_{(n)}(\beta,\overline\mu)$ in the set $\mathscr G_n$ do not come from $f_*$.
To get the asymptotic properties of the estimators defined above, we introduce the following conditions.
\begin{itemize}\item[{\it C1}.] The intercept of the model (\ref{(2.1)}) is zero, $\varepsilon$ is independent of $X$, $E[XX^T]$ is a positive definite matrix, and the variances $\sigma_i^2$ of $\varepsilon_i$  with distribution $f_i$ exist for all $i=1,\cdots, N$.
\item[{\it C2}.] The distribution $f_*$ satisfying (\ref{(2.5)}) is unique and the size $n_*$ of the sample from $f_*$ tends to infinity as $N\rightarrow\infty$.
\item[{\it C3}.] $\lambda=n^{\epsilon-1}$ for a constant $0<\epsilon<1$.
\item[{\it C4}.] In the case of $n\geq n_0$, $(n-n_*)/n^{1-\epsilon}=o(1)$, $n^{1-\epsilon}/n_0<C$ for a constant $C>0$, and $\Delta_{n_d}=o(n^{1-\varepsilon})$,
where $n_d=n-d_n$.
\end{itemize}

For the above four conditions, we have the following explanations.
The first two conditions in {\it C1} are standard. 
Condition {\it C2} is based on (\ref{(2.4)}) and (\ref{(2.5)}). This condition implies that the second-order moment condition $E_{f_*}[(\varepsilon-\overline\mu)^2]>E_{f}[(\varepsilon-\overline\mu)^2]$ for all $f\neq f_*,f\in\mathscr F$. Based on this condition, we can judge whether the corresponding errors $\varepsilon_{k_j},j=1,\cdots,n_*,$ come from the same distribution $f_*$. The use of the uniqueness assumption on $f_*$ in {\it C2} is to get a simple estimation procedure. However, this uniqueness assumption may not be always true. Thus, it will be removed when an adjusted method is introduced in the next section.
We need condition {\it C3} to constrain the convergence rate at which $\lambda \Delta_n$ tends to zero. The first two conditions in {\it C4} are based on the fact that most of $G_{(j)}(\beta,\overline\mu),j=1,\cdots,n$, come from the distribution $f_{*}$ (see Subsection 2.1). The two conditions give the range of $n$ when the penalized estimation is used.
The third condition in {\it C4} is also standard. 

Denote $\mu_*=E_{f_*}[\varepsilon]$, $\sigma_*^2=E_{f_*}[(\varepsilon-\overline\mu)^2]$ and
$\Phi(X)=
\left(\begin{array}{ccc}XX^T&X\\X^T&1\end{array}\right).$ We have the following theroem.

\begin{theorem}\label{theorem3.1} {\it Under the model (\ref{(2.1)}), suppose  conditions {\it C1}-{\it C4} hold.
Then the PMLS estimator defined in (\ref{(3.4)}) satisfies
$$\sqrt{n_*}\left[(\widehat\beta-\beta)^T,\widehat{\overline\mu}-\mu_*\right]^T\stackrel{ d}\longrightarrow N\left(0,\sigma_*^2E^{-1}[\Phi(X)]\right)\ \ (n_*\rightarrow\infty),$$ where $\stackrel{d}\longrightarrow$ stands for convergence in distribution.
}\end{theorem}

The proof of the theorem is given in Appendix. The key of the proof is to show that most of the elements in $\mathscr G_n$ come from $f_*$ via the penalty in (\ref{(3.4)}). The proof implies that the uniqueness assumption on $f_*$ is unnecessary. In the next section, the assumption can be removed via an additional penalty.

The theorem guarantees that the PMLS estimator $\widehat\beta$ is consistent and normally distributed asymptotically. However, the PMLS estimator $\widehat{\overline\mu}$ is not always consistent because it tends to $\mu_{*}$, rather than the true parameter $\overline \mu$.
On the other hand, compared with the properties of parameter estimation in the case of classical nonlinear regression, here the variance is enlarged and the convergence rate is reduced to $1/\sqrt {n_*}$. This is mainly because of the variability of the error terms, which comes from  distribution uncertainty. 

\subsection{Second-step estimator of $\overline \mu$}

We now go to the second-step of the estimation to refine the estimator of $\overline\mu$ to achieve the consistency. Similar to (\ref{(2.4)}) and (\ref{(2.5)}), suppose the following holds:\begin{eqnarray}\label{(3.5)}\overline\mu=\mathbb E[\varepsilon]=\sup_{f\in\mathscr F} E_f[\varepsilon]=E_{\widetilde f}[\varepsilon] \ \ \mbox{ for a }\ \widetilde f\in \mathscr F.\end{eqnarray} Let  $\{H_{(j)}=Y_{s_j}-\widehat\beta^TX_{s_j}:j=1,\cdots,N\}$ be the order statistics of $\{H_j=Y_{j}-\widehat\beta^TX_{j}:j=1,\cdots,N\}$ in descending order
$H_{(1)}\geq H_{(2)}\geq \cdots \geq H_{(N)}.$
Similar to the decomposition in (\ref{(3.1)}), the index set $I_n=\{s_j:j=1,\cdots,n\}$
is decomposed as $I_n=U_n\cup L_n$.
Then, by the same argument as used in the first-step estimation, the second-step estimator of $\overline\mu$ is defined by
\begin{eqnarray}\label{(3.6)}\widehat{\overline\mu}_{Sec}=\arg\min_{\overline\mu \in\mathscr U,n_{\widetilde\lambda}\in\mathscr N}\frac {1}{n_{\widetilde\lambda}}\sum_{j=1}^{n_{\widetilde\lambda}}\left(H_{(j)}-\overline\mu\right)^2+\widetilde\lambda\, |\Gamma_{n_{\widetilde\lambda}}|,\end{eqnarray}  where
$$\Gamma_{n}=\frac {1}{[n/2]}\sum\limits_{j\in U_n}\left(Y_{j}-\widehat\beta^TX_{j}\right)-\frac {1}{n-[n/2]}\sum\limits_{j\in L_n}\left(Y_{j}-\widehat\beta^TX_{j}\right).$$ Here the tuning parameter $\widetilde\lambda\geq 0$ may be different from that in (\ref{(3.4)}), but also satisfies condition {\it C3}. As argued above,  the objective function in (\ref{(3.6)}) is a convex function of $\overline \mu$. The estimator of (\ref{(3.6)}) is a PMLS estimator as well. Comparing with the estimation procedure in (\ref{(3.4)}), the data set $\{(X_{s_j},Y_{s_j}):j=1,\cdots,\widetilde n\}$ used here should be different from the data set $\{(X_{k_j},Y_{k_j}):j=1,\cdots, n_*\}$ used in (\ref{(3.4)}).

Let $\widetilde n$ be the size of the sample from $\widetilde f$. The following conditions are required to establish the estimation consistency for the second-step estimator of $\overline\mu$.
\begin{itemize}
\item[{\it C5}.]  $\widetilde n\rightarrow\infty$ and $n_*/{\widetilde n}\rightarrow c\neq 0$ as $N\rightarrow\infty$.
    \item[{\it C6}.] Condition {\it C4} holds when the notations are replaced by the above accordingly.
\end{itemize} Unlike {\it C2}, here the uniqueness assumption on $\widetilde f$ is not required. It is because the penalty for $\Gamma_{ n}$ in (\ref{(3.6)}) can make sure that most of $\varepsilon_{s_j},j=1,\cdots,\widetilde n$, have the common mean $\overline\mu$.

\begin{theorem}\label{theorem3.2} {\it Under the conditions in Theorem 3.1, conditions {\it C5} and {\it C6}, when  $\widetilde\lambda$ satisfies the same condition of $\lambda$ as given in  condition {\it C3}, and  $\{\varepsilon_{k_j},j=1,\cdots,n_*\}$ and $\{\varepsilon_{s_j},j=1,\cdots,\widetilde n\}$ are not overlapped, then the second-step estimator in (\ref{(3.6)}) satisfies
$$\sqrt{\widetilde n}\left(\widehat{\overline\mu}_{Sec}-\overline\mu\right)\stackrel{ d}\longrightarrow N\left(0,\widetilde\sigma^{2}+\sigma_*^2E[X^T]E[\Omega^{-1}(X)]E[X]\right)\ \ (\widetilde n\rightarrow\infty),$$ where
$\widetilde\sigma^{2}=E_{\widetilde f}[(\varepsilon-\overline\mu)^2]$ and
\begin{eqnarray*}\Omega^{-1}(X,\theta)=(E[XX^T])^{-1}+
(E[XX^T])^{-1}E[X]E[X^T](E[XX^T])^{-1}/c\end{eqnarray*} with $c=1-E[X^T](E[XX^T])^{-1}E[X].$
}
\end{theorem}

The proof of the theorem is presented in Appendix. Here we need the constraint of non-overlapping between  $\{\varepsilon_{k_j},j=1,\cdots,n_*\}$ and $\{\varepsilon_{s_j},j=1,\cdots,\widetilde n\}$ only for the simplicity of proof and representation. This condition can be replaced by $f_*\neq \tilde f$ and can be further reduced to that the number $n^o$ of overlapping elements in these two sets $\{\varepsilon_{s_j},j=1,\cdots,\widetilde n\}$ and $\{\varepsilon_{k_j},j=1,\cdots,n_*\}$ satisfies $n^o/\widetilde n=o(1)$. After $n_*$ and $\widetilde n$ being determined, the condition can be checked by the methods of testing distributions to be equal; the details are omitted here. By the theorem, the second-step PMLS estimator $\widehat{\overline\mu}_{Sec}$ is consistent and normally distributed asymptotically.

\subsection{A summary of the algorithm}

The above estimation procedures involves four tuning parameters: $\lambda$, $\widetilde\lambda$, $n_\lambda$ and $n_{\widetilde\lambda}$. These parameters can be chosen by the cross-validation that is similar to those used in variable selection for
high-dimensional models (see, e.g., Fan and Li (2001)). Since the set of the tuning parameters contains $n_\lambda$ and $n_{\widetilde\lambda}$, the numbers of data used for constructing the estimators, the cross-validation used here should be somewhat different from that given in Fan and Li (2001). If the discrete function $s(n)=\frac 1n\sum_{j=1}^nG_{(j)}(\beta,\overline\mu)$ is approximated by a continuously differentiable one, and a prior distribution $\pi(\beta,\overline\mu,n)$ for $(\beta,\overline\mu,n)$ is assumed, then, by the Bayesian information criterion (see Schwarz (1978)), the Bayesian cross-validations for $\theta=(\lambda,n_\lambda)$ and $\widetilde \theta=(\widetilde \lambda,n_{\widetilde\lambda})$ can be defined respectively as
$$CV(\theta)=CV(\theta)+\frac{(p+2)\log n_\lambda}{n_\lambda}, \ \ CV(\widetilde \theta)=CV(\widetilde \theta)+\frac{(p+2)\log n_{\widetilde\lambda}}{n_{\widetilde\lambda}},$$ where $CV(\cdot)$ is the cross-validation criterion defined by Fan and Li (2001). The above Bayesian cross-validations do not depend on the prior distribution, and they are in fact the large-sample criteria beyond the Bayesian context.
Combining the  above estimation procedure with the cross-validation for tuning parameter selection, the whole algorithm can be summarized into the following steps:

{\it Step 1. Initial estimator of  $(\beta,\overline\mu)$.} Let $(\beta^1,\overline \mu^1)$ be an initial selection of $(\beta,\overline\mu)$, and $\{(Y_{k_j}-X_{k_j}^T\beta^1-\overline\mu^1)^2:j=1,\cdots,N\}$ be the order quantities of the original squared quantities $\left\{(Y_{j}-X_{j}^T\beta^1-\overline\mu^1)^2:i=1,\cdots,N\right\}$ in  descending order. For each pair of the tuning parameters $\theta=(\lambda,n_\lambda)$, the full data set $T=\{(X_{k_j},Y_{k_j}):i=1,\cdots,n_\lambda\}$ is first divided at random into cross-validation training sets $T-T^\nu$ and test sets $T^\nu$, $\nu=1,\cdots,5$, and then the initial estimator $\left(\widehat{\beta}^{(\nu)}(\theta),\widehat{\overline\mu}^{(\nu)}(\theta)\right)$ is obtained by the training set $T-T^\nu$ via (\ref{(3.4)}).

{\it Step 2. Selection of $\theta$.}  Write $G_{i}^{(\nu)}(\theta)
=\left(Y_{i}-X_{i}^T\widehat{\beta}^{(\nu)}(\theta)
-\widehat{\overline\mu}^{(\nu)}(\theta)\right)^2$ and
let $\left\{G_{(j)}^{(\nu)}(\theta)
=\left(Y_{k_j}-X_{k_j}^T\widehat{\beta}^{(\nu)}(\theta)
-\widehat{\overline\mu}^{(\nu)}(\theta)\right)^2:(X_{k_j},Y_{k_j})\in T^\nu\right\}$ be the order statistic of $\left\{G_{i}^{(\nu)}(\theta):(X_{k_j},Y_{k_j})\in T^\nu\right\}$ in the descending order. Define a Baysian cross-validation criterion as
$$CV\left(\theta\right)=\frac {1}{n_\lambda}\sum_{\nu=1}^5\sum_{(X_{k_j},Y_{k_j})\in T^\nu,1\leq j\leq n_\lambda}G_{(j)}^{(\nu)}\left(\theta\right)+\frac{(p+2)\log n_\lambda}{n_\lambda}.$$ We then get an estimator $\widehat\theta$ by minimizing $CV\left(\theta\right)$.

{\it Step 3. Final estimator of $\beta$.} With the selected estimator $\widehat\theta$, we estimate $\beta$ as the first component $\widehat\beta$ of the following vector:
\begin{eqnarray}\label{(3.7)}\left(\widehat\beta^T,\widehat{\overline\mu}\right)^T=\arg\min_{\beta\in\mathscr B,\overline\mu\in\mathscr U}\frac {1}{\widehat n_\lambda}\sum_{j=1}^{\widehat n_\lambda}G_{(j)}(\beta,\overline\mu)+\hat\lambda\left|\Upsilon^2_{\widehat n_\lambda}(Y,\overline\mu)-2\overline X^T\beta\Upsilon^1_{\widehat n_\lambda}(X,Y)\right|.\end{eqnarray}

{\it Step 4. Initial estimator of $\overline \mu$.} With the estimator $\widehat\beta$ obtained above, and for each pair of the tuning parameters $\widetilde\theta=(\widetilde\lambda,n_{\widetilde\lambda})$ and the training set $T-T^\nu$, we find the estimator $\widehat{\overline\mu}^{(\nu)}(\widetilde\theta)$ by (\ref{(3.6)}).

{\it Step 5. Selection of $\widetilde\theta$.}  Write $G_{i}^{(\nu)}(\widetilde\theta)
=\left(Y_{i}-X_{i}^T\widehat{\beta}
-\widehat{\overline\mu}^{(\nu)}(\widetilde\theta)\right)^2$ and
let $$\left\{G_{(j)}^{(\nu)}(\widetilde\theta)
=\left(Y_{k_j}-X_{k_j}^T\widehat{\beta}
-\widehat{\overline\mu}^{(\nu)}(\widetilde\theta)\right)^2:(X_{k_j},Y_{k_j})\in T^\nu\right\}$$ be the order statistic of $\left\{G_{i}^{(\nu)}(\widetilde\theta):(X_{k_j},Y_{k_j})\in T^\nu\right\}$, satisfying $G_{(1)}^{(\nu)}(\widetilde\theta)\geq G_{(2)}^{(\nu)}(\widetilde\theta)\geq\cdots \geq G_{(n_\lambda)}^{(\nu)}(\widetilde\theta)$. Define the Bayesian cross-validation criterion as
$$CV(\widetilde\theta)=\frac {1}{n_{\widetilde\lambda}}\sum_{\nu=1}^5\sum_{(X_{k_j},Y_{k_j})\in T^\nu,1\leq j\leq n_{\widetilde\lambda}}G_{(j)}^{(\nu)}(\widetilde\theta)+\frac{2\log n_{\widetilde\lambda}}{n_{\widetilde\lambda}}.$$ We then get an estimator $\widehat{\widetilde\theta}$ by minimizing $CV(\widetilde\theta)$.

{\it Step 6. Final estimator of $\overline \mu$.} With the selected estimator $\widehat{\widetilde\theta}$, we
estimate $\overline\mu$ by
\begin{eqnarray}\label{(3.8)}\widehat{\overline\mu}_{Sec}=\arg\min_{\overline\mu \in\mathscr U}\frac {1}{\widehat n_{\widetilde\lambda}}\sum_{j=1}^{\widehat n_{\widetilde\lambda}}\left(H_{(j)}-\overline\mu\right)^2+\widehat{\widetilde\lambda} \,|\Gamma_{\widehat n_{\widetilde\lambda}}|.\end{eqnarray}

Alternatively, we can use $\tau$ instead of $\lambda$ as the tuning parameter and together with the others to design the algorithm. As the method is similar to the above and thus the details are omitted.

\setcounter{equation}{0}
\section{Extension and discussion}

It is known that there may be more than one distribution in $  \mathscr F $ that can attain the upper expectation. In this section we first recommend an extended method to remove the uniqueness assumption on $f_*$ in {\it C2}. It can be seen from the proof of Theorem 3.1 that  the uniqueness assumption is only to guarantee that  most of $\varepsilon_{k_j},j=1,\cdots,n_*$, have the same mean $\mu_*$. Then the uniqueness assumption on the distributions $f_*$ can be removed. All we need to do in the data selection procedure is to identify the data that satisfy the first two order moment conditions: $\varepsilon_{k_j},j=1,\cdots,n_*$, have the same mean $\mu_*$ and the variance $\sigma_*^2$.

Let $\{D_{(j)}=Y_{l_j}-X^T_{l_j}\widehat\beta_{LS},j=1,\cdots,N\}$ be the order statistics of $\{D_j=Y_{j}-X^T_{j}\widehat\beta_{LS},j=1,\cdots,N\}$, satisfying
$D_{(1)}\geq D_{(2)}\geq\cdots \geq D_{(N)}.$ Similar to (\ref{(3.1)}), for the index set $I_n=\left\{l_j:j=1,\cdots,n\right\}$, the decomposition is designed as $I_n=U_n\cup L_n$.
Write
$$\Lambda_{n}(X,Y)=\frac {1}{[n/2]}\sum\limits_{j\in L_n}D_{j}-\frac {1}{n-[n/2]}\sum\limits_{j\in L_n}D_{j}.$$
The proof of Lemma 3.1 given in Appendix  shows
$$\frac 1n\sum_{i=1}^n(Y_i-X^T_i\widehat\beta_{LS})= \frac 1n\sum_{i=1}^n\mu_i-\frac 1N\sum_{j=1}^N\mu_jE[X^T]E^{-1}
[XX^T]E[X]+O_p\left(\frac{1}{\sqrt{N}}\right).$$
Thus, we can use $\Lambda_{n}(X,Y)$ to measure the difference among the means $\mu_{l_j},j=1,\cdots,n$, and then use $$|\Lambda_{n}(X,Y)|<\tau_1$$ to control the difference among the means $\mu_{l_j}$ for all $j$. Consequently,
an improved estimator of $\beta$ is defined as the first component of the following solution:
\begin{eqnarray}\label{(4.1)}&&\nonumber \hspace{-6ex}\left(\widehat\beta^T_I,\widehat{\overline\mu}_I
\right)^T\\&&\hspace{-6ex}=\arg\min_{\beta\in\mathscr B,\overline\mu\in\mathscr U,n_\lambda\in\mathscr N}\Big\{\frac 1{n_\lambda}\sum_{j=1}^{n_\lambda}G_{(j)}(\beta,\overline\mu) \\ &&\nonumber \ \ \ \ \ \ \ \ \ \ \ \ +\lambda \Big|\Upsilon^2_{ n_\lambda}(Y,\overline\mu)-2\overline X^T\beta\Upsilon^1_{ n_\lambda}(X,Y)\Big|+\lambda_1|\Lambda_{n_\lambda}(X,Y)|\Big\},\end{eqnarray} where $\lambda\geq 0$ and $\lambda_1\geq0$ are two tuning parameters.
We now use two penalties $\Lambda_{n}$ and $\Upsilon^2_{n}(Y,\overline\mu)-2\overline X^T\beta\Upsilon^1_{n}(X,Y)$ to make sure that the selected data satisfy the first and second order moment conditions. Here a possible choice of $n_\lambda$ is
$$n_\tau=\max\left\{n:|\Upsilon^2_{n}(Y,\overline\mu)-2\overline X^T\beta\Upsilon^1_{n}(X,Y)|<\tau,|\Lambda_{n}(X,Y)|<\tau_1\right\}.$$

Without the uniqueness assumption on $f_*$, the condition
{\it C2} is replaced by
\begin{itemize}
\item[{\it C7}.] The number $n_c$ of the errors satisfying the first two order moment conditions tends to infinity as $N\rightarrow\infty$.
\end{itemize} Now we do not need the uniqueness assumption on the distribution, however, the size of the data set used here is usually smaller than that  in the previous section. We have the following theorem.

\begin{theorem}\label{Theorem 4.1}
{\it Under the model (\ref{(2.1)}), suppose  conditions {\it C1}, {\it C4} and {\it C7} hold, $\lambda$ and $\lambda_1$ satisfy condition {\it C3}.
Then the PMLS estimator defined in (\ref{(4.1)}) satisfies
$$\sqrt{n_c}\left[(\widehat\beta_I-\beta)^T,\widehat{\overline\mu}_I-\mu_*\right]^T\stackrel{ d}\longrightarrow N\left(0,\sigma_*^2E^{-1}[\Phi(X)]\right)\ \ (n_c\rightarrow\infty),$$ where $\mu_*$ and $\sigma_*^2$ are defined in the previous section.
}\end{theorem}

The proof of the theorem is given in Appendix. As stated above, the uniqueness assumption on $f_*$ is removed from the estimation procedure at the cost that  the convergence rate of the improved estimator can be slower than that of $(\widehat\beta,\widehat{\overline\mu})$, but the asymptotic normality still holds.

Also we can use the second-step estimation procedure as given in the previous section to construct the consistent estimator for $\overline\mu$. Let  $\{H_{(j)}^I=Y_{m_j}-\widehat\beta^T_IX_{m_j}:j=1,\cdots,N\}$ be the order statistic of $\{H_j^I=Y_{j}-\widehat\beta^T_I X_{j}:j=1,\cdots,N\}$, satisfying
$H_{(1)}^I\geq H_{(2)}^I\geq \cdots \geq H_{(N)}^I.$ Denoted by $I_n=U_n\cup L_n$ the decomposition of the index set $I_n=\{m_j:
j=1,\cdots,n\}$ as in (\ref{(3.1)}). Then, by the same argument as used above, the second-step estimator of $\overline\mu$ is defined by
\begin{eqnarray}\label{(4.2)}\widehat{\overline\mu}_{Sec}^I=\arg\min_{\overline\mu \in\mathscr U,n_{\widetilde\lambda}\in\mathscr N}\frac {1}{n_{\widetilde\lambda}}\sum_{j=1}^{n_{\widetilde\lambda}}\left(H_{(j)}^I-\overline\mu\right)^2
+
\widetilde\lambda\, |\Gamma^I_{n_{\widetilde\lambda}}|,\end{eqnarray} where
$$\Gamma^I_{n}=\frac {1}{[n/2]}\sum\limits_{j\in U_n}H_{j}^I-\frac {1}{{n}-[n/2]}\sum\limits_{j\in L_n}
H_{j}^I.$$ Then, this second-step estimator is consistent. The following theorem states the result.

\begin{theorem}\label{Theorem 4.2} {\it Under the conditions of Theorem 4.1, conditions {\it C5} and {\it C6}, when $\widetilde\lambda$ satisfies  condition {\it C3} and $\{\varepsilon_{l_j},j=1,\cdots,n_c\}$ and $\{\varepsilon_{m_j},j=1,\cdots,\widetilde n\}$ are not overlapped, then the second-step estimator in (\ref{(4.2)}) satisfies
$$\sqrt{\widetilde n}\left(\widehat{\overline\mu}_{Sec}^I-\overline\mu\right)\stackrel{ d}\longrightarrow N\left(0,\widetilde\sigma^{2}+\sigma_*^2E[X^T]E[\Omega^{-1}(X)]E[X]\right)\ \ (\widetilde n\rightarrow\infty),$$ where
$\widetilde\sigma^{2}$, $\sigma_*^2$ and $\Omega(X,\theta)$ are defined in the previous section.
}\end{theorem}

The difficulty we are facing now is the computational complexity because there are five tuning parameters: $\lambda$, $\lambda_1$, $\widetilde\lambda$, $n_\lambda$ and $ n_{\widetilde\lambda}$. The computational steps are similar to those in the previous section. Because of the complexity, if we have the prior information on the uniqueness of the distribution $f_*$, we prefer to use the method given in the previous section to construct the estimators.

We now discuss the special case of $\beta=0$. In this case, the model is simplified as
\begin{eqnarray}\label{(4.3)}Y=\varepsilon.\end{eqnarray} We can see how the upper expectation $\overline\mu=\mathbb E[\varepsilon]=\mathbb E[Y]$ is estimated consistently whereas existing result only derives
$\underline\mu\leq\overline Y \leq \overline\mu$ as we mentioned in Section~1. Although the methods proposed above can be used, for this simple model, estimation can be much simpler. Let $\{Y_{(j)}=Y_{t_j},j=1,\cdots, N\}$ be the order statistics of $\{Y_{j},j=1,\cdots, N\}$ with descending order $Y_{(1)}\geq Y_{(2)}\geq \cdots \geq Y_{(N)}$.
For the index set $I_n=\{t_j:j=1,\cdots,n\}$, we define the decomposition as $I_n=U_n\cup L_n$ as (\ref{(3.1)}). Write
$$\Delta_n(Y)=\frac {1}{[n/2]}\sum\limits_{j\in U_n}Y_{j}-\frac {1}{n-[n/2]}\sum\limits_{j\in L_n}Y_{j}.$$ Then, the estimator for $\overline\mu$ is defined by
\begin{eqnarray}\label{(4.4)}\widehat{\overline\mu}=\arg\min_{\overline\mu\in\mathscr U,n_\lambda\in\mathscr N}\frac {1}{n_\lambda}\sum_{j=1}^{n_\lambda}(Y_{(j)}-\overline\mu)^2 +\lambda \Delta_{n_\lambda}(Y),\end{eqnarray} where $\lambda\geq0$ is a tuning parameter as well.  

Let $\widetilde n$ be the sample size from $\widetilde f$ given in (\ref{(3.5)}). We need the following simpler conditions than before:

\begin{itemize} \item[{\it C8}.]
The variances $\sigma_i^2$ of $\varepsilon_i$ exist for all $i=1,\cdots, N$.
\item[{\it C9}.]   $\widetilde n\rightarrow\infty$ as $N\rightarrow\infty$.
    \item[{\it C10}.] Condition {\it C4} holds when the notations are replaced by the above accordingly.
\end{itemize}

\begin{theorem}\label{Theorem4.3}
{\it Suppose  that conditions {\it C8}-{\it C10} hold. Then the PMLS estimator $\widehat{\overline\mu}$ defined in (\ref{(4.4)}) satisfies
$$\sqrt{\widetilde n}(\widehat{\overline\mu}-\overline\mu)\stackrel{ d}\longrightarrow N\left(0,\sigma_*^2\right)\ \ (\widetilde n\rightarrow\infty).$$
}\end{theorem}

Before ending this section, we briefly discuss how to extend the methods proposed above to the nonlinear regression function $g(\beta,X)$ as the estimation procedure is almost the same. We thus omit the detail. But, as the least squares estimation requires the derivative of $g(\beta,X)$ with respect to the parameter $\beta$, 
we need to assume that  $$\frac{\partial g(\beta,X)}{\partial\beta_j}\not\equiv 0 \ \mbox{ for } j=1,\cdots,p,$$ are the non-constant functions of $X$. This condition is to guarantee the identifiability of the regression function $g(\beta,X)$. For example, in the regression model:
$$Y=\beta_1+\exp{(-\beta_2-\beta_3^X)}+\varepsilon,$$ the parameter $\beta_1$ is unidentifiable as it cannot be separated out from $\theta_0+\overline\mu$. 

\setcounter{equation}{0}
\section{Numerical studies}

\subsection{Simulation studies} In this subsection we examine the finite sample behaviors of the newly proposed estimators by simulation studies. To obtain thorough
comparisons, in addition to the new PMLS estimator, we
comprehensively consider several competitors such as the OLS estimators that ignore the distribution uncertainty.
Mean
squared error (MSE), prediction error (PE) and boxplots are used to evaluate the performances of the involved
estimators and models. Also the simulation results for
estimation bias are reported to emphasize the influence from distribution uncertainty, especially from  expectation uncertainty.  In the following, we design 4 experiments. The first experiment is to compare the PMLS with the overall average  $\overline Y$ for estimating the upper expectation of $Y$, the second and third experiments are designed for examining the performances of PMLS and OLS when estimating the parameter $\beta$ and $\overline \mu$ in simple linear and multiple linear models. The fourth experiment is used to investigate the usefulness of PMLS for prediction.

{\it Experiment 1.} Consider the  simplest case with $\beta=0$:
$$Y_i=\varepsilon_i,\ \ i=1,\cdots, N,$$ where $\varepsilon_i,i=1,\cdots, N,$ are independent and follow the distributions in the class $\mathscr F=\left\{N(\mu,\sigma^2):(\mu,\sigma^2)\in{\cal T}\right\}$. In the following, we respectively consider two cases of distribution uncertainty:
\begin{itemize}\item[] { \it Case 1}. ${\cal T}=\left\{k/2:k=1,\cdots,10\right\}\times\left\{0.20^2,0.25^2\right\}$ and $T=(\mu,\sigma^2)$ is uniformly distributed on ${\cal T}$.
\item[] {\it Case 2}. ${\cal T}=\left\{k:k=1,\cdots,10\right\}\times\left\{0.25^2\right\}$ and $T=(\mu,\sigma^2)$ is uniformly distributed on ${\cal T}$.\end{itemize}
For each $k$, the size of the sample from $N(\mu_k,\sigma_k^2)$ is designed as $[N/10]$.

Before performing the simulation, we first use the histograms of $Y_i$ in the two cases to observe what pattern of the data appears to show distribution uncertainty. It is very clear from Figures 1 and 2 that the distributions in the two cases are multimodal although every distribution is unimodal.
It shows that when we have a data set showing multimodal pattern, we may not simply believe the multimodality of an underlying distribution, distribution uncertainty would also be a possibility.  Under this situation, the classical statistical inferences such as the estimation of population expectation, have less accuracy. Instead,
our goal is to consistently estimate the upper expectation $\overline\mu=\mathbb E[\varepsilon]=\mathbb E[Y]$.

\begin{center} Figure 1 about here\end{center}

\begin{center} Figure 2 about here\end{center}

To examine the consequence of ignoring distribution uncertainty in estimation, we compare the PMLS estimator with the OLS estimator that is the overall average $\overline Y$ of all of observations in this experiment.
In the simulation procedure, for each $k$, the size of the sample from $N(\mu_k,\sigma_k^2)$ is designed as $[N/10]$.
For the total sample sizes $N=100,500$ and $1000$, the empirical bias and MSE, and the boxplots of the estimators with 500 replications are reported respectively in Tables~1 and 2, and Figures~3 and 4. Note that for this very simple model, we cannot have a constant intercept term because of distribution uncertainty. Therefore, theoretically, the intercept term for every observation is not identifiable, which is absorbed in the error term in the upper expectation of error term. The OLS estimator estimates nothing as its limit is in between, from the description in Section~1,  the upper and lower expectation: $\underline \mu \le \bar Y \le\overline \mu$ with a probability going to one.
The simulation results can verify that our PMLS estimator is clearly superior to the OLS estimator. More precisely, we have the following findings:
\begin{itemize}\item[] (1)  From Tables~1 and 2, distribution uncertainty mainly results in the estimation bias of the OLS estimator, and the estimation bias almost obliterates the effect of variance in the MSE of the estimator. However, the  uncertainty has no significant impact for the PMLS estimator for the upper expectation $\overline \mu$.  The estimation bias  of the PMLS estimator are very obviously smaller than those of the OLS estimator in both cases. The centerlines of the boxplots of the PMLS estimator are just located respectively at the true values 5 and 10 of the upper expectations. But the centerlines of the boxplots of the OLS estimator are far below the true values.  
\item[] (2) From Figures~3 and 4, we can see that although the boxplots of the PMLS estimator are nearly centralized around the centerlines, the values have more dispersion than  those of the OLS estimator, implying the new estimator has larger variance and a slow convergence rate. It is because the new method only uses a part of the data. However, this enlarged variance is negligible compared with the significant estimation bias of which the OLS estimator suffers. \end{itemize}

\begin{center} Table 1 about here\end{center}

\begin{center} Figure 3 about here\end{center}

\begin{center} Table 2 about here\end{center}

\begin{center} Figure 4 about here\end{center}

{\it Experiment 2.} Consider the following univariate linear regression:
$$Y_i=\beta X_i+\varepsilon_i,\ \ i=1,\cdots, N,$$ where  $X_i,i=1,\cdots, N$, are independent and identically distributed as $N(1,1)$. Suppose that $\varepsilon_i,i=1,\cdots, N,$ are independent and follow the distributions in the class $\mathscr F=\left\{N(\mu,\sigma^2):(\mu,\sigma^2)\in{\cal T}\right\}$ with ${\cal T}=\{k:k=1,\cdots,10\}\times\{(0.05k)^2:k=1,\cdots,10\}$. As we commented in Section~2,  the model cannot contain a nonzero constant intercept term because even an intercept term is imposed, it is impossible to be identified  and consistently estimated.  The histograms of $Y_i$ and the residuals $\hat\varepsilon_i$ derived from the OLS are
multimodal to present distribution uncertainty (the histograms are not reported herewith for saving space).

In the simulation, we set $\beta=2$ and let
$(\mu,\sigma^2)$ be uniformly distributed on ${\cal T}$. For each $k$, the size of the sample from $N(\mu_k,\sigma_k^2)$ is designed as $[N/10]$.
For total sample sizes $N=100,500$ and $1000$, the empirical bias and MSE, and the boxplots of the estimators over 500 replications are reported respectively in Table 3, and Figures 5 and 6. Although the model used here is totally different from that in Experiment 1, a conclusion from the simulation results is similar to the finding (1) obtained in Experiment 1. That is to say, PMLS  can accurately estimate the regression coefficient and the upper expectation of the error, while  OLS gets the estimators that are far away from the true values. Unlike that in the finding (2) in Experiment~1, the variance of the OLS estimator is larger than that of the PMLS estimator in this experiment. The PMLS estimator of $\beta$ performs  better than the  PMLS estimator of $\overline\mu$ with smaller bias and MSE particularly when the sample size is large. Perhaps it is because the two-step estimation procedure for $\overline\mu$  introduces more estimation error.

We note that the OLS estimator $\widehat\beta_{LS}$ has a significant bias. One may expect to centralize data to reduce the bias. As we explained before, for every observation,  the center $E_{f_{t_i}}(\varepsilon_i)$ is a conditional expectation when $T=t_i$ is given, and such a center is actually a random variable because of the distribution uncertainty defined in Section 2. Thus, in theory,  using the overall average of $Y_i$'s as the center of every $Y_i$ is not meaningful and it is also not estimable. On the other hand, in practice, when we use it as if distribution uncertainty did not exist, its practical performance can be promoted because $\varepsilon_i$ is not centered. We now pretend that the observations do not have distribution uncertainty.  If only $Y_i$'s in the above regression are centered, it can be easily verified that, in this example, $$\widehat\beta_{LS}=\frac 12 \beta+O_p\left(\frac{1}{\sqrt{N}}\right),$$ which is also biased. If both $X_i$' and $Y_i$' are centered, it can be seen that $$\widehat\beta_{LS}=\beta+O_p\left(\frac{1}{\sqrt{N}}\right).$$  The simulation result in Table 4 shows that when we blindly use OLS with centered data,  the estimation efficiency does be promoted. However, even for the latter, the estimation bias is slightly larger than that of the PMLS estimator, and  the MSE of the centered LS estimator is about 5 times of that of the the PMLS estimator
although in the case without distribution uncertainty, the bias-reduction LS estimator should have a variance  achieving Fisher information bound.

\begin{center} Table 3 about here\end{center}

\begin{center} Figure 5 about here\end{center}

\begin{center} Figure 6 about here\end{center}

\begin{center} Table 4 about here\end{center}

{\it Experiment 3.} Consider the  multiple  linear regression:
$$Y_i=\beta_1 X_{1i}+\beta_2 X_{2i}+\varepsilon_i,\ \ i=1,\cdots, N,$$ where $\beta_1=3$, $\beta_2=2$, $X_{1i}\sim N(1,1)$ and $X_{2i}\sim N(2,1)$, the other settings are designed as those  in Experiment 2. The simulation results are reported in Table~5 and Figures 7-9 and further indicate that the PMLS estimator is consistently superior to the OLS estimator in estimation bias, MSE and variance.  Again we can see that  OLS overestimates regression coefficients because the distribution uncertainty makes it impossible to remove the bias which is absorbed in the error terms. The effect of the overestimated regression coefficients by  OLS is to compensate the loss of ignoring the positive error. Then, a new problem emerges naturally:  {\it Is  OLS  able to give a proper prediction?} We  discuss this issue in the following experiment.

\begin{center} Table 5 about here\end{center}

\begin{center} Figure 7 about here\end{center}

\begin{center} Figure 8 about here\end{center}

\begin{center} Figure 9 about here\end{center}

{\it Experiment 4.} The model and experiment conditions are completely identical to those in Experiment~3, but the purpose is to examine the prediction behavior. Before comparing the predictions derived by  OLS and PMLS, we define the meaning of prediction under the situation with distribution uncertainty. Because the classical methods ignore  distribution uncertainty, a natural prediction of $Y$ based on OLS is given as
$$\widehat Y_{LS}=\widehat\beta_{LS}^TX_0$$ for a given predictor $X_0$. However, the main goal of the upper expectation regression is to predict maximum values of $Y$ conditional on predictor $X_0$. Thus, under the framework of upper expectation regression, the prediction is defined by
$$\widehat Y_M=\widehat\beta^TX_0+\widehat{\overline\mu},$$ where both $\widehat\beta$ and  $\widehat{\overline\mu}$ are the MPLS estimators proposed in the previous sections. On the other hand, if our goal is to predict all the values $Y_i$, not merely the maximum values, based on PMLS, a reasonable prediction is defined as
$$\widehat Y_{PMLS}=\widehat\beta^TX_0+\mu_M,$$ where $\mu_M$ is a suitable value in the interval $[\widehat{\underline\mu},\widehat{\overline\mu}]$. Here the estimator $\widehat{\underline\mu}$ of the lower expectation of $\varepsilon$ can be obtained by the similar argument proposed in the previous sections. If without additional information about  expectation uncertainty of $\varepsilon$, we simply choose the middle point $\mu_M=(\widehat{\underline\mu}+\widehat{\overline\mu})/2$. It is worth pointing out that although the overall average of $\overline Y-\widehat\beta^T\overline X$ can also be between $\widehat{\underline\mu}$ and $\widehat{\overline\mu}$, it does not converge to a fixed value under distribution uncertainty and thus, its use makes no theoretical ground.

We first consider the performances of the predictions for some larger values of $Y$.
For $n$ values $Y_1,\cdots,Y_n$, we rearrange them in  descending order as
$$Y_{(1)}\geq Y_{(2)}\geq \cdots\geq Y_{(n)}.$$ In this case,
the average prediction error (APE) of predicting the first $m$ largest values of $Y$ is defined by
$$APE=\frac 1 m\sum_{i=1}^m(Y_{(i)}-\widehat Y_{(i)})^2,$$ where $\widehat Y_{(i)}$ is a prediction value of $Y_{(i)}$.
The simulation results are presented in Figure 10, in which the curves are the medians of APEs of 500 replications. It clearly shows that the upper expectation regression can relatively accurately predict the larger values of $Y$. More precisely, for 100 values of $Y$, the upper expectation regression gives relatively successful prediction for the first $34$ largest values of $Y$. However, if ignoring distribution uncertainty, the OLS-based prediction behaves poorly for predicting the larger values of $Y$.

\begin{center} Figure 10 about here\end{center}

Finally, we investigate the behaviors of the predictions $\widehat Y_{LS}$ and $\widehat Y_{PMLS}$ for all the values of $Y$. The APEs of the two predictions are reported in Table 6. It is clear that the APE of $\widehat Y_{PMLS}$ is significantly smaller than that of $\widehat Y_{LS}$. Because of distribution uncertainty, however, both the two predictions have relatively large APE even for large sample size. It shows that it is impossible to improve the predictions
if without further information about the distribution of $\varepsilon$, in other words, we can not completely characterize the regression under the situation with distribution uncertainty.

\begin{center} Table 6 about here\end{center}

\subsection{Real data analysis}

In this subsection we use a real data example to show how the upper expectation regression works under the setting with distribution uncertainty. We consider the data set of the Fifth National Bank of Springfield based on data
from 1995 (see examples 11.3 and 11.4 in Albright et al., 1999). This data set has been analyzed  such as Fan and Peng (2004) and Cui et al. (2013). The bank, whose name has since
changed, was charged in court with paying its female employees substantially lower salaries
than its male employees. For each of its 208 employees, the data set includes the following
variables:

\begin{itemize} \item EduLev: education level, a categorical variable with categories 1 (finished
high school), 2 (finished some college courses), 3 (obtained a bachelor¡¯s
degree), 4 (took some graduate courses), 5 (obtained a graduate degree).

\item JobGrade: a categorical variable indicating the current job level, the possible
levels being 1-6 (6 highest).

\item  YrHired: year that an employee was hired.

\item  YrBorn: year that an employee was born.

\item  Gender: a categorical variable with values ``Female" and ``Male".

\item  YrsPrior: number of years of work experience at another bank prior to
working at the Fifth National Bank.

\item PCJob: a dummy variable with value 1 if the empolyee's current job is
computer related and value 0 otherwise.

\item  Salary: current (1995) annual salary in thousands of dollars.\end{itemize}

\noindent   Fan and Peng (2004) employed the  linear model as
\begin{eqnarray}\label{(5.1)}\mbox{Salary}= \beta_0 +\beta_1 \mbox{Gender} +\beta_2 \mbox{PCJob}+
\sum_{i=1}^4\beta_{2+i}\mbox{Edu}_i+
\sum_{i=1}^5\beta_{6+i}\mbox{JobGrd}_i +\varepsilon.\end{eqnarray}
We first use histogram of salary to examine data distribution. We can see from  Figure~11 that the histogram of Salary exhibits multimodality. We do not simply consider the error  to have a multimodal distribution. This is because other factors, such as the years of working experience and the age of  an employee, may affect the salary. Therefore, we regard these potential factors as latent factors in an upper expectation regression with distribution uncertainty.  Consider the upper expectation linear model to fit the data: \begin{eqnarray}\label{(5.3)} \mathbb E(\mbox{Salary})= \beta_1 \mbox{Gender} +\beta_2 \mbox{PCJob}+
\sum_{i=1}^4\beta_{2+i}\mbox{Edu}_i+
\sum_{i+1}^5\beta_{6+i}\mbox{JobGrd}_i +\overline\mu.\end{eqnarray}
It is worth pointing out that the differences from the model (\ref{(5.1)}) are that the error $\varepsilon$ in (\ref{(5.3)}) is supposed to be of distribution uncertainty, and the model (\ref{(5.3)}) does not have  the intercept term, which is included in the upper expectation of $\varepsilon$. 

\begin{center} Figure 11 about here\end{center}


We use 170 data to estimate the model parameters and then use the obtained models to fit the rest of the data (to predict 38 values of ``Salary"). The predictions and prediction errors are defined in Experiment 4. The results of parameter estimation and the APE are listed in Table 7. Compared with the OLS regression (\ref{(5.1)}), the upper expectation regression (\ref{(5.3)}) has the following interesting features:

\begin{itemize}\item[(1)] The absolute values of the estimators of the  coefficients of JobGrd$_i$ are significantly reduced, but the others, especially  the coefficients of Gender and Edu$_2$, are largened. We may explain these as follows. As  JobGrd$_i$ may be related to the years of working experience and the age of the employee, when these factors are not included in model (\ref{(5.1)}), the model requires larger coefficients of JobGrd$_i$ to draw the information of these factors. On the other hand,  the effect of JobGrd$_i$ is absorbed into the error of model~(\ref{(5.3)}).
 \item[(2)] The difference of the APEs between the two models is not significant.\end{itemize}

\begin{center} Table 7 about here\end{center}

\noindent On the other hand, as shown in Experiment 4, the upper expectation regression is more concerned about the maximum information. Figure 13 presents the medians of the APEs for the $m$ largest values of ``Salary" via 100 replications. From this figure, we can get the following finding:

\begin{itemize} \item[(3)] The upper expectation regression can relatively accurately predict the larger values of ``Salary". For example, for the first 15 largest values of ``Salary", the APE of the upper expectation regression is 10.6892 smaller than that of the OLS regression; and for the first 24 largest values of  of ``Salary", the APE of the upper expectation regression is 6.5338 smaller than that of the OLS regression.\end{itemize}

\noindent Finally, we examine the $R^2$ values of the two models, which is defined by
$$R^2_m=1-\frac{\sum_{j=1}^m(Y_{(j)}-\widehat Y_{(j)})^2}{\sum_{j=1}^m(Y_{(j)}-\overline Y_m)^2}\ \ \mbox{ for } m\leq N,$$ where $\overline Y_m=\sum_{j=1}^mY_{(j)}/m$ with $Y_{(j)}$ given in Experiment 4. We use the values of $R^2_m$ of the first $m$ largest values of ``Salary" to check if the upper expectation regression can capture the maximum risk information. The result is reported in Figure 14. It indicates the following conclusion:
\begin{itemize} \item[(4)] For the two models, most values of $R^2$  are larger than 0.79, while the upper expectation regression has a relatively high $R^2$ for the larger values of ``Salary".\end{itemize}
All the numerical results aforementioned are coincident with the theoretical conclusions.

\begin{center} Figure 12 about here\end{center}

\begin{center} Figure 13 about here\end{center}


\setcounter{equation}{0}
\section{Appendix: Proofs}


\noindent{\it Proof of (\ref{(2.7)}).} It is clear that
$$P\left(\bigcup_{k=m}^nA_k\right)\leq\sum_{k=m}^nP(A_k)=\sum_{k=m}^nC_n^kp^k(1-p)^{n-k}.$$ By the definition of $p$ and a simple integral operation, we can get
$$p=\frac 34-\frac{1}{n+1}+\frac{1}{(n+1)4^{n+1}}.$$ It followings from this result and the stirling's formula that
$$C_n^kp^k(1-p)^{n-k}\sim \frac{n^n}{k^k(n-k)^{n-k}}\frac{3^k}{4^n}.$$ It can be verified that $\frac{n^n}{k^k(n-k)^{n-k}}\frac{3^k}{4^n}$ is a decreasing function of $k$ when $k$ is large enough. Then
$$\sum_{k=m}^nC_n^kp^k(1-p)^{n-k}\sim (n-m)\frac{n^n}{m^m(n-m)^{n-m}}\frac{3^m}{4^n}.$$ Note that $m=[n^{\delta}]$ for $0<\delta<1$. Then
$$(n-m)\frac{n^n}{m^m(n-m)^{n-m}}\frac{3^m}{4^n}\rightarrow 0,$$ implying the result of (\ref{(2.7)}). $\Box$

\noindent {\it Proof of Lemma 3.1.} For simplicity we only consider the case of $n>n_*$ and $n=2m$.

It can be easily proved that, under the assumption {\it C1},
$$\widehat\beta_{LS}-\beta=\frac 1N\sum_{j=1}^N\mu_jE^{-1}
[XX^T]E[X]+O_p\left(\frac{1}{\sqrt{N}}\right).$$
The result leads to
$$Y_i-X^T_i\widehat\beta_{LS}= \varepsilon_i-\frac 1N\sum_{j=1}^N\mu_jX^T_iE^{-1}
[XX^T]E[X]+O_p\left(\frac{1}{\sqrt{N}}\right),$$ implying
$$\frac 1n\sum_{i=1}^n(Y_i-X^T_i\widehat\beta_{LS})= \frac 1n\sum_{i=1}^n\mu_i-\frac 1N\sum_{j=1}^N\mu_j
E[X^T]E^{-1}
[XX^T]E[X]+O_p\left(\frac{1}{\sqrt{n}}\right).$$
Then,
under model (\ref{(2.1)}), if $\varepsilon$ is independent of $X$, then
\begin{eqnarray*}\nonumber\hspace{-6ex}&&\frac 1n\sum_{j=1}^n(Y_{j}-\overline\mu)^2
\\\nonumber\hspace{-6ex}&&=\frac 1n\sum_{j=1}^n(X_{j}^T\beta+\varepsilon_{j}-\overline\mu)^2\\&&=C_1+\frac 1n\sum_{j=1}^nE[G_{j}(\beta,\overline\mu)]+
\frac 2n\sum_{j=1}^n\mu_{j} E[X^T]\beta+O_p\left(\frac{1}{\sqrt{n}}\right)\\\nonumber&&=C_N+\frac 1n\sum_{j=1}^nE[G_{j}(\beta,\overline\mu)]+
\frac 2n\sum_{j=1}^n(Y_{j}-X^T_{j}\widehat\beta_{LS})E[X^T]\beta+O_p\left(\frac{1}{\sqrt{n}}\right)
\\\nonumber&&=C_N+\frac 1n\sum_{j=1}^nE[G_{j}(\beta,\overline\mu)]+
\frac 2n\sum_{j=1}^n(Y_{j}-X^T_{j}\widehat\beta_{LS})\overline X^T\beta+O_p\left(\frac{1}{\sqrt{n}}\right),
\end{eqnarray*} where $C_1=\beta^T E[X X^T]\beta-2\overline\mu E[X^T]\beta$ and $$C_N=C_1+\frac 1N\sum_{j=1}^N\mu_jE[X^T]E^{-1}
[XX^T]E[X]E[X^T]\beta.$$
The relation above leads to the conclusion of the lemma. $\Box$

\noindent{\it Proof of Theorem 3.1.}
It can be see from Lemma 3.1 and (\ref{(a.7)}) given below that for the asymptotic property of parameter estimation, the objective functions in (\ref{(3.3)}) and (\ref{(3.4)}) respectively have the following equivalent forms:
$$\frac 12\gamma' V\gamma + U_{n_\tau}\gamma\ \mbox{ and } \ \frac 12\gamma' V\gamma + U_{n_\lambda}\gamma+r_{n_\lambda}(\gamma),$$ where $\gamma$ is a parameter vector, $V$ is a positive definite matrix, $U_n$ is stochastically bounded and $r_n(\gamma)$ goes to zero in probability for each $\gamma$. Thus, by the basic corollary of Hj\o rt and Pollard (1993), we have that the objective functions in (\ref{(3.3)}) and (\ref{(3.4)}) are equivalent for parameter estimation with respect to asymptotic property. We thus only investigate the asymptotic properties of the estimator defined by (\ref{(3.3)}).

For simplicity, here we only consider the case when $n$ is a even number: $n=2m$. Note that $\frac {1}{n}\sum\limits_{j=1}^{n}G_{(j)}(\beta,\overline\mu)$ and $\Delta_n$ are respectively decreasing and increasing functions of $n$ when $n$ exceeds $n_0$. This leads to that the selected $n$ should satisfy $n\geq n_0$. In this case, it can be verified that $\Delta_n\geq 0$. Suppose without loss of generality that, for $n\geq n_0$, only the last $d_n$ elements  $G_{(n-d_n+1)}(\beta,\overline\mu),\cdots, G_{(n)}(\beta,\overline\mu)$ in the set $\mathscr G_n$ do not come from $f_*$, with $n-d_n$ being an even number: $n-d_n=2k$. Then,  \begin{eqnarray*}\Delta_n&=&\frac{1}{n}\left(\sum_{j=1}^kE[G_{(j)}(\beta,\overline\mu)]-\sum_{j=1}^kE[G_{(k+j)}
(\beta,\overline\mu)]\right)\\&&+\frac 2n\sum_{j=1}^{m-k}E[G_{(k+j)}(\beta,\overline\mu)]-\frac 1n\sum_{j=1}^{d_n}E[G_{(2k+j)}(\beta,\overline\mu)]\\&=:&\frac 1n I_1+\frac 1n I_2-\frac 1n I_3.\end{eqnarray*} By the treatments above, we have
$$\lambda\Delta_n=\frac {1}{n^{1-\epsilon}} I_1+\frac {1}{n^{1-\epsilon}} I_2-\frac {1}{n^{1-\epsilon}} I_3.$$ By the above result and the condition {\it C4}, it is clear that if $n=O(n_0)$, then $\lambda\Delta_n=o(1)$. If $n/n_0$ is diverging, then $(n_*+d_n)/n_0\rightarrow\infty$, implying $d_n/n_0\rightarrow\infty$. Note that $n^{1-\epsilon}/n_0$ is bounded and $$\frac{d_n}{n_0}=\frac{d_n}{n^{1-\epsilon}}\frac{n^{1-\epsilon}}{n_0}.$$ Thus
 $\frac{d_n}{n^{1-\epsilon}}\rightarrow\infty,$ resulting in $\frac {1}{n^{1-\epsilon}} I_2\rightarrow\infty$ and $\frac {1}{n^{1-\epsilon}} I_3\rightarrow\infty$.
In this case, $\lambda\Delta_n$ is diverging as well and, consequently, the minimum value of the objective function $\frac {1}{n}\sum\limits_{j=1}^{n}G_{(j)}(\beta,\overline\mu)+\lambda\Delta_n$ does not exist.
We then need only to consider the objective function $\frac 1{n}\sum_{j=1}^{n}G_{(j)}(\beta,\overline\mu)+\lambda|\Delta_n|$ with $n=O(n_0)$ for the asymptotic properties of the estimation.

Furthermore, because $n=O(n_0)$, the objective function can be further expressed as
\begin{eqnarray}\label{(a.4)}\frac 1n\sum\limits_{j=1}^{n-d_n}G_{(j)}(\beta,\overline\mu)+o_p(1)=
\frac 1n\sum\limits_{j=1}^{n_*}G_{(j)}(\beta,\overline\mu)+o_p(1).\end{eqnarray}
Denoted by $\beta^0$ and $\mu^0_*$ the true values of $\beta$ and $\mu_*$, respectively, and let $\beta$ and $\overline\mu$ satisfy  $\|\beta-\beta^0\|=O(1/\sqrt n)$ and $|\overline\mu-\mu_{*}^0|=O(1/\sqrt n)$. Because $n=O(n_0)$, we can assume $n_*/n\rightarrow 1$, without loss of generality. Then, the objective function (\ref{(a.4)}) can be replaced by
\begin{eqnarray}\label{(a.5)}\nonumber&&\frac 1{n_*}\sum\limits_{j=1}^{n_*}G_{(j)}(\beta,\overline\mu)+o_p(1)\\\nonumber&&
=\frac{1}{n_*}\sum_{j=1}^{n_*}
\left(\varepsilon_{k_j}+{\beta^0}'X_{k_j}-{\beta}'X_{k_j}-\overline\mu\right)^2+o_p(1)
\\\nonumber&&=\frac{1}{n_*}\sum_{j=1}^{n_*}
\left\{(\varepsilon_{k_j}-\mu_{*}^0)^2-2[(\beta-\beta^0)'X_{k_j}+(\overline\mu-\mu_{*}^0)]
(\varepsilon_{k_j}-\mu_{*}^0)\right.\\&&\left.\hspace{2.5cm}  +[(\beta-\beta^0)'X_{k_j}+(\overline\mu-\mu_{*}^0)]^2\right\}+o_p(1)\end{eqnarray} Because $\sum_{j=1}^{n_*}
(\varepsilon_{k_j}-\mu_{*}^0)^2$ is free of $\beta$ and $\overline\mu$, the objective function in (\ref{(a.5)}) is equivalent to
\begin{eqnarray}\label{(a.6)}\nonumber&&\frac{1}{n_*}\sum_{j=1}^{n_*}
\left\{-2[(\beta-\beta^0)'X_{k_j}+(\overline\mu-\mu_{*}^0)]
(\varepsilon_{k_j}-\mu_{*}^0)\right.\\&&\left.\hspace{2.5cm} +[(\beta-\beta^0)'X_{k_j}+(\overline\mu-\mu_{*}^0)]^2\right\}+o_p(1).\end{eqnarray} By the basic corollary of Hj\o rt and Pollard (1993), the term of order $o_p(1)$ can be ignored for the asymptotic property of the estimation. We then rewrite the above objective function as
\begin{eqnarray}\label{(a.7)}Z_n(\gamma)=\sum_{j=1}^{n_*}\left\{\frac{-2}{\sqrt{n_*}}[\varepsilon_{k_j}-\mu_{*}^0]
[X_{k_j}',1]\gamma+\frac{1}{n_*}\gamma'\Phi(X_{k_j})\gamma\right\}.\end{eqnarray}
The objective function $Z_n(\gamma)$ is obviously convex and is minimized at
$$\Gamma_n=\sqrt{n_*}[(\widehat\beta-\beta^0)',\widehat{\overline\mu}-\mu_{*}^0]'.$$ Note that $\varepsilon_{k_j},j=1,\cdots,n_*$, are identically distributed with the common mean $\mu_{*}^0$ by the condition {\it C2}. It follows from the Lindeberg-Feller central limit theorem that
$$Z_n(\gamma)\stackrel{d}\longrightarrow Z_0(\gamma)=-2W'\gamma+\gamma'E[\Phi(X)]\gamma,$$ where $W\sim N(0,u^2_{*}E[\Phi(X)])$. The convexity of the limiting objective function, $Z_0(\gamma)$, assures the uniqueness of the minimizer and consequently, that
$$\sqrt{n_*}\left[(\widehat\beta-\beta^0)',\widehat{\overline\mu}-\mu_{*}^0\right]'=\hat\gamma_n=\arg\min\widetilde Z_n(\gamma)\stackrel{d}\longrightarrow\hat\gamma_0=\arg\min Z_0(\gamma).$$
(See, e.g., Pollard 1991, Hj\o rt and Pollard 1993, Knight 1998). Finally, we see $\hat\gamma_0=E^{-1}[\Phi(X)]W$ and $(n_*)/n\rightarrow 1$. Then the result follows.
$\Box$

\noindent{\it Proof of Theorem 3.2.} By the same argument as used in the proof of Theorem 3.1, $n_{\widetilde \tau}$ can be replaced by the sample size $\widetilde n$. Let $\{H^0_{(j)}=Y_{s_j}-\beta^TX_{s_j}:j=1,\cdots,N\}$ be the order statistic of $\{ H^0_j=Y_{j}-\beta^TX_{j}:j=1,\cdots,N\}$, satisfying $H^0_{(1)}\geq H^0_{(2)}\geq\cdots\geq H^0_{(n)}$.
Write the corresponding index decomposition as $C^0_n=U^0_n\cup L^0_n$ and let
$$\Gamma^0_{n_{\widetilde\lambda}}=\frac {1}{[n_{\widetilde\lambda}/2]}\sum\limits_{j\in U^0_n}H^0_{j}-\frac {1}{\widetilde n-[n_{\widetilde\lambda}/2]}\sum\limits_{j\in L^0_n} H^0_{j}.$$ It follows from Theorem 3.1 that $\Gamma_{n_{\widetilde\lambda}}=\Gamma^0_{n_{\widetilde\lambda}}+O_p(1/\sqrt{n_{\widetilde\lambda}})$.
Denoted by $\beta^0$ and $\overline\mu^0$ the true values of $\beta$ and $\overline\mu$, respectively, and  let $\beta$ and $\overline\mu$ satisfy $\|\beta-\beta^0\|=O(1/\sqrt {n_{\widetilde\lambda}})$ and $|\overline\mu-\overline\mu^0|=O(1/\sqrt {n_{\widetilde\lambda}})$. Thus,
the objective function in (\ref{(3.6)}) can be expressed as
\begin{eqnarray*}\frac {1}{n_{\widetilde\lambda}}\sum_{j=1}^{n_{\widetilde\lambda}}\left([\varepsilon_{s_j}-\overline\mu^0]-(\beta^0-\widehat\beta)^TX_{s_j}
-[\overline\mu-\overline\mu^0]\right)^2-\widetilde\lambda \Gamma_{n_{\widetilde\lambda}}+O_p(1/\sqrt{n_{\widetilde\lambda}}).\end{eqnarray*} Note that $\{\varepsilon_{k_j},j=1,\cdots,n_*\}$ and $\{\varepsilon_{s_j},j=1,\cdots,n_{\widetilde\lambda}\}$ are independent, and $\widehat\beta$ depends only on $\{\varepsilon_{k_j},j=1,\cdots,n_*\}$. By the conclusion of Theorem 3.1 and the same argument as used in the proof of Theorem 3.1, we can prove the theorem. $\Box$

\noindent{\it Proofs of Theorem 4.1 - 4.3.} The proofs are similar to that of Theorem 3.1. $\Box$

\

\newpage

\begin{figure}
\begin{center}
\begin{tabular}[b]{c}
{\includegraphics[width=9cm,height=7cm]{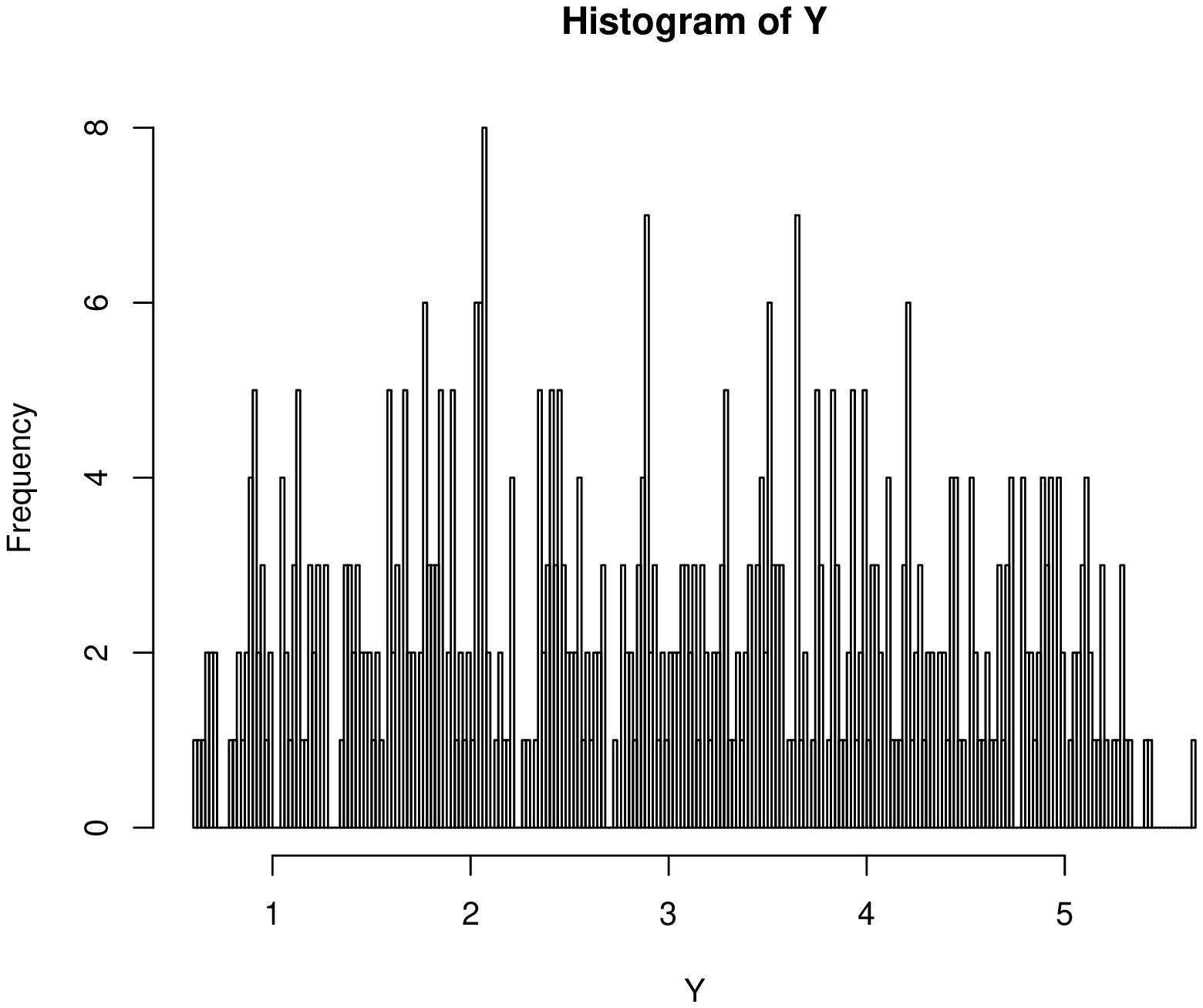}}
\end{tabular}
\vspace{-4ex}\caption{Histogram for case 1 with $N=500$.} \label{fig:1}
\end{center}
\end{figure}

\begin{figure}
\begin{center}
\begin{tabular}[b]{c}
{\includegraphics[width=9cm,height=7cm]{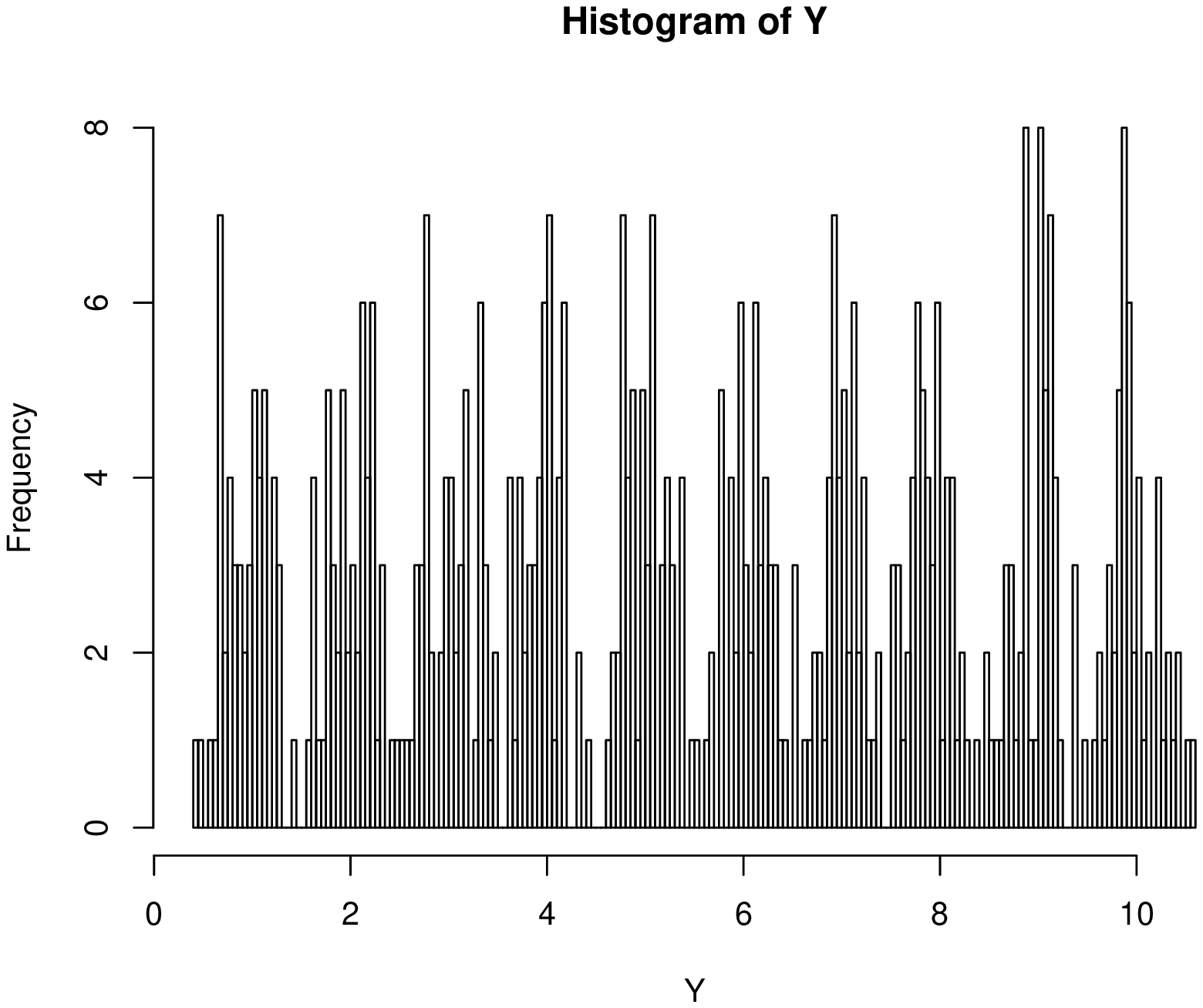}}
\end{tabular}
\vspace{-4ex}\caption{Histogram for case 2 with $N=500$.} \label{fig:2}
\end{center}
\end{figure}

\clearpage

\begin{table}
\caption{Estimation bias and MSE for  case 1 in Experiment 1}
\label{tab:1} \vspace{0.3cm} \center
\begin{tabular}{c||cccccc}
  \hline
$N$ & 100& & 500 & &1000& \\
criterions& Bias& MSE & Bias& MSE & Bias& MSE \\\hline
 PMLS & $-0.0842$ &0.0319 & $0.0640$& 0.0113&0.1210& 0.0187  \\
 OLS &$-2.2517$ & 5.0707 & $-2.2501$& 5.0631& $-2.2493$&5.0595\\
  \hline
\end{tabular}
\end{table}

\begin{figure}
\begin{minipage}{4cm}
\begin{tabular}[b]{c}\ \ \ \ \ \
{\includegraphics[width=4.5cm,height=6cm]{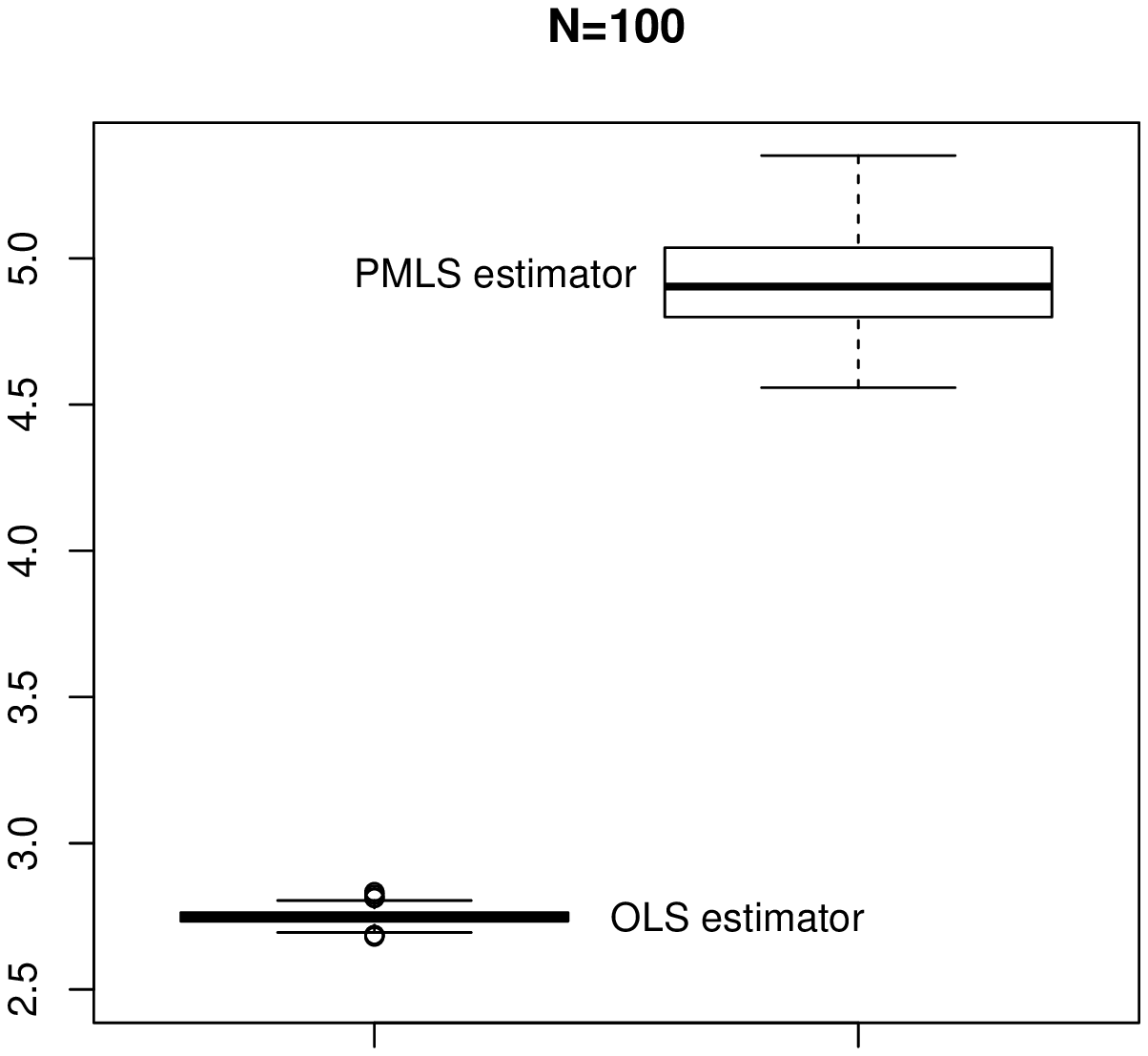}}
\end{tabular}
\end{minipage}
\begin{minipage}{4cm}
\begin{tabular}[b]{c}\ \ \ \ \ \
{\includegraphics[width=4.5cm,height=6.3cm]{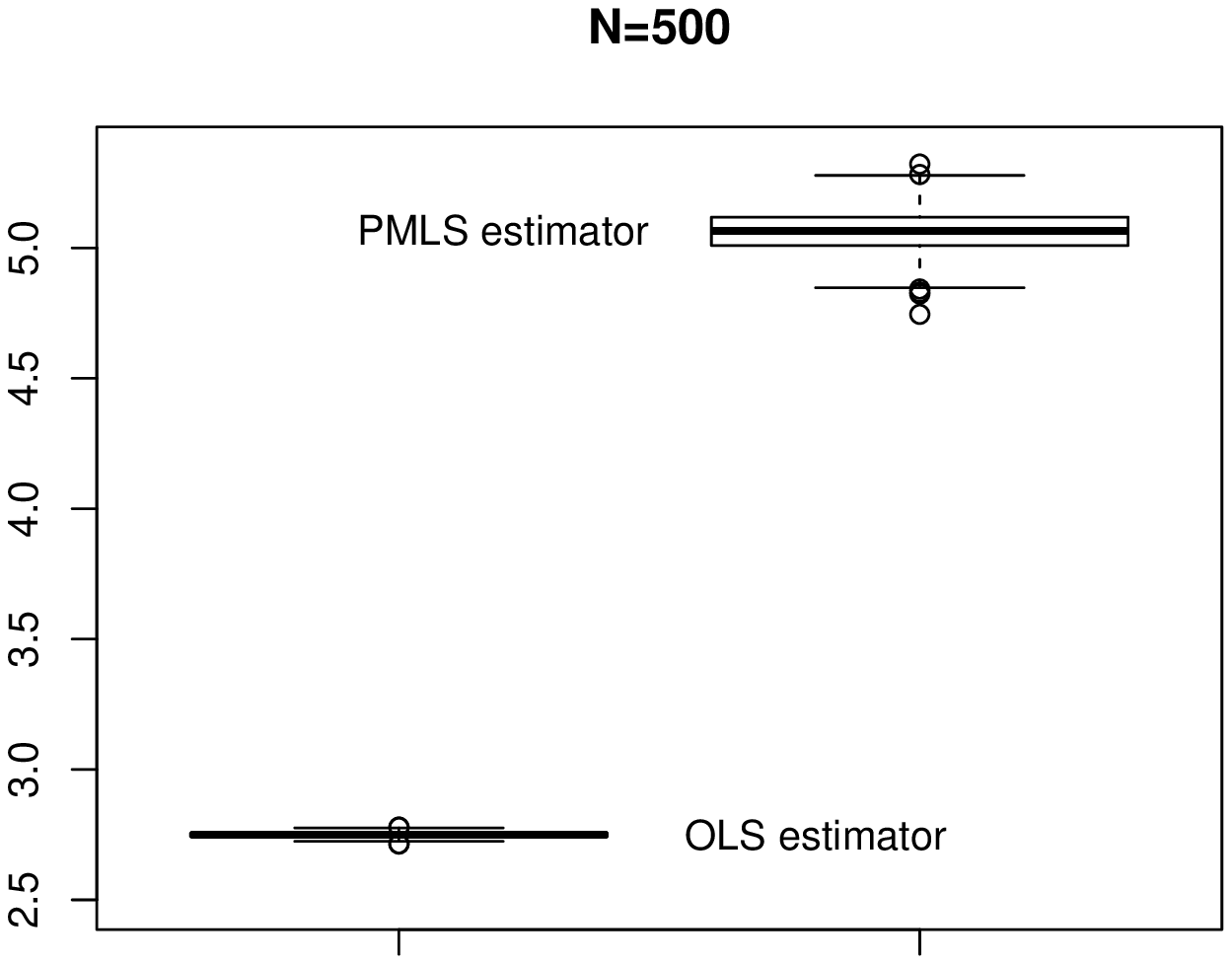}}
\end{tabular}
\end{minipage}
\begin{minipage}{4cm}
\begin{tabular}[b]{c}\ \ \ \ \ \
{\includegraphics[width=4.5cm,height=6.6cm]{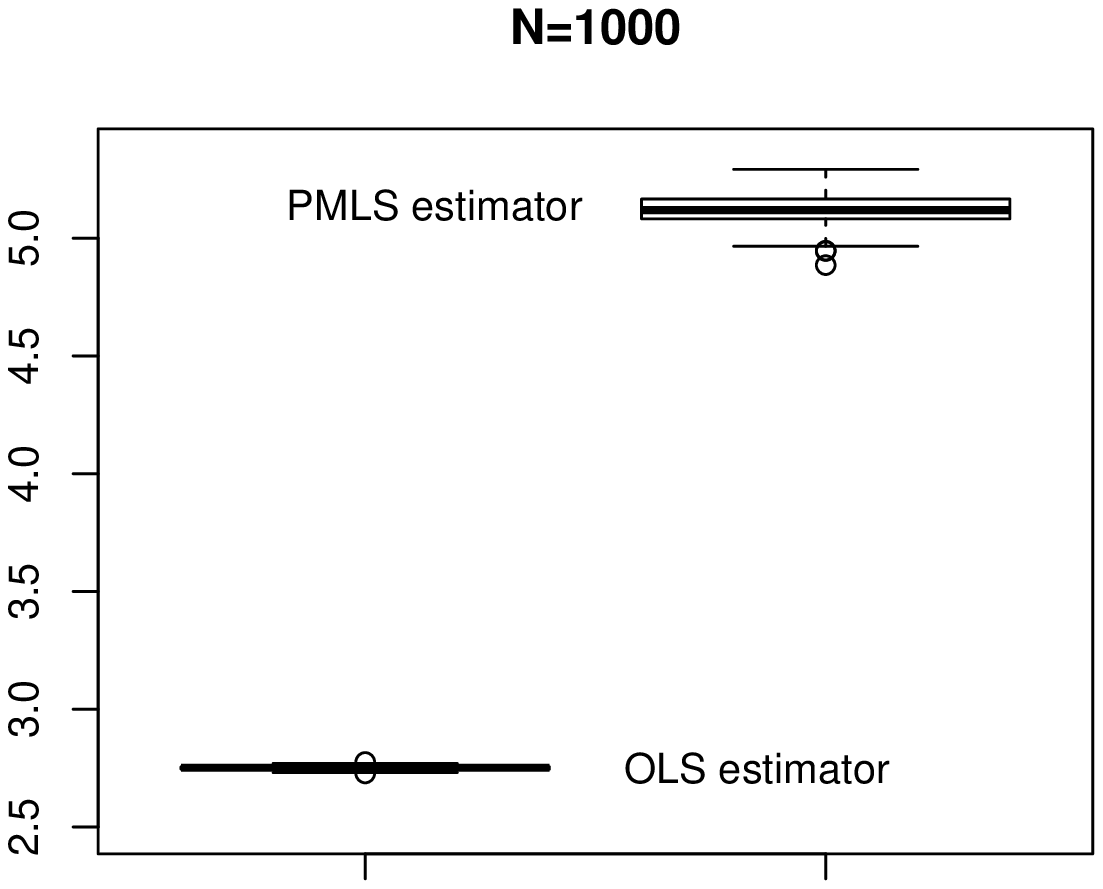}}
\end{tabular}
\end{minipage}\vspace{-6ex}
\caption{The boxplots of the PMLS  estimator and the OLS estimator in case 1 with the true $\overline\mu=5$ in Experiment 1.} \label{fig:3}
\end{figure}

\begin{table}
\caption{Estimation bias and MSE for case 2 in Experiment 1}
\label{tab:2} \vspace{0.3cm} \center
\begin{tabular}{l||cccccc}
  \hline
$N$ & 100& & 500 & &1000& \\
 criterions& Bias& MSE & Bias& MSE & Bias& MSE \\\hline
 PMLS & $-0.0408$ &0.0157 & $0.0562$& 0.0062&0.0920& 0.0108  \\
 OLS &$-4.4980$ & 20.2330 & $-4.4999$& 20.2497& $-4.5000$&20.2509\\
  \hline
\end{tabular}
\end{table}

\begin{figure}
\begin{minipage}{4cm}
\begin{tabular}[b]{c}\ \
{\includegraphics[width=5cm,height=7.5cm]{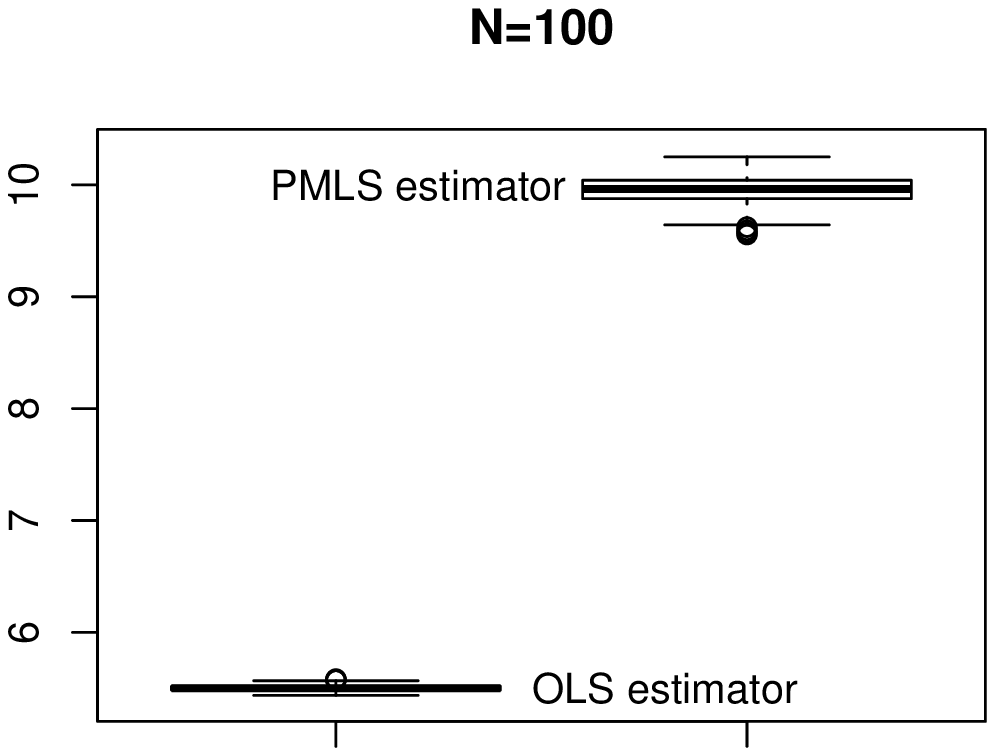}}
\end{tabular}
\end{minipage}
\begin{minipage}{4cm}
\begin{tabular}[b]{c}\ \
{\includegraphics[width=5cm,height=7.5cm]{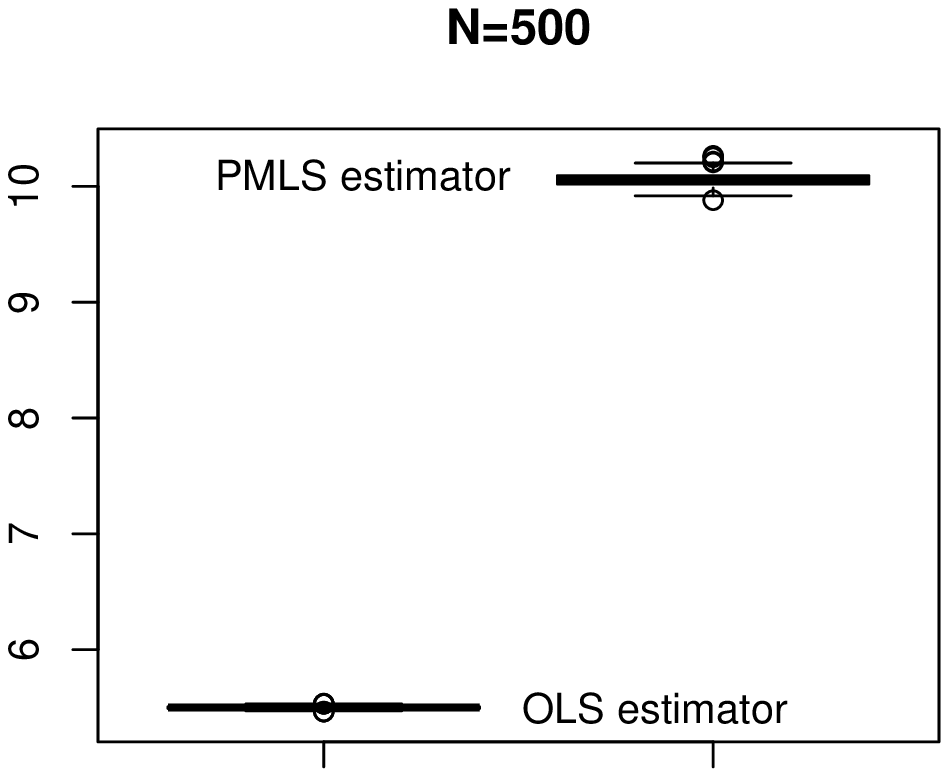}}
\end{tabular}
\end{minipage}
\begin{minipage}{4cm}
\begin{tabular}[b]{c}\ \
{\includegraphics[width=5cm,height=7.5cm]{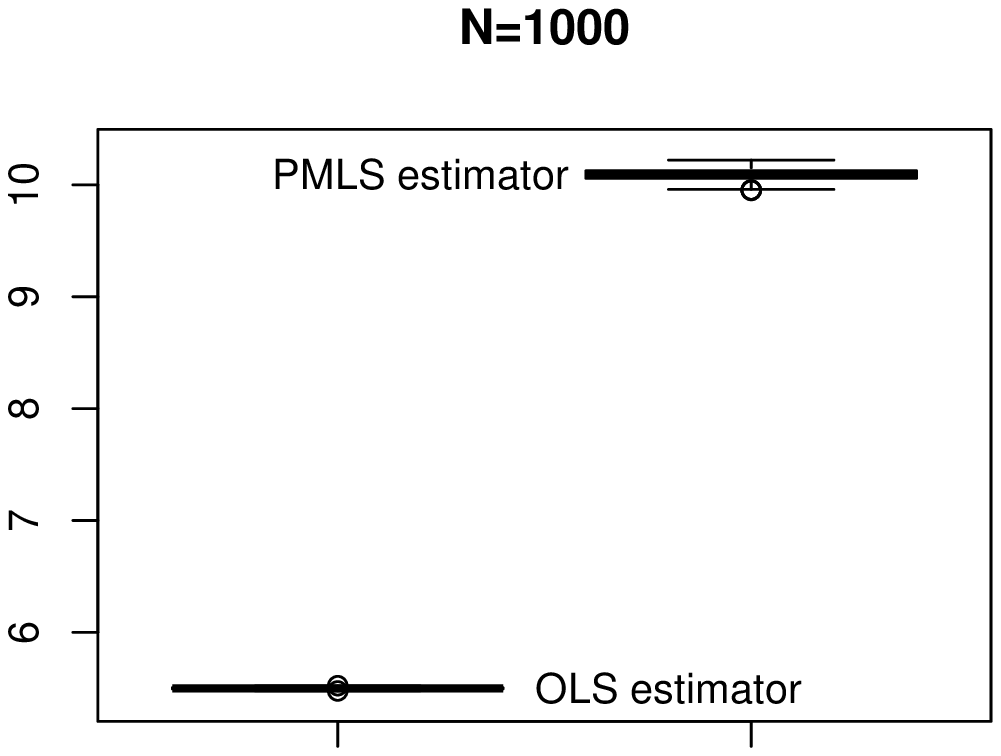}}
\end{tabular}
\end{minipage}\vspace{-6ex}
\caption{The boxplots of the PMLS estimator and the OLS estimator in case 2 with the true $\overline\mu=10$ in Experiment 1.} \label{fig:4}
\end{figure}

\newpage

\

\

\

\

\

\begin{table}
\caption{Estimation bias and MSE in Experiment 2}
\label{tab:3} \vspace{0.3cm} \center
\begin{tabular}{c||ccccccc}
  \hline
 parameters & $N$ & 100& & 500 & &1000& \\
& criterions& Bias& MSE & Bias& MSE & Bias& MSE \\\hline
$\beta$ & PMLS & $-0.07497$ &0.03101 & $-0.01854$& 0.00341&$-0.00449$& 0.00063  \\
& OLS &$2.75808$ &7.66282 & $2.74548$& 7.55041& $2.74865$&7.56090\\
  \hline
$\overline\mu$   & PMLS & $0.00636$ &0.13711 & $0.16631$& 0.05203&0.26512& 0.08386\\
  \hline
\end{tabular}
\end{table}

\begin{table}
\caption{Estimation bias and MSE for the bias-reduced LS estimator}
\label{tab:4} \vspace{0.3cm} \center
\begin{tabular}{c||cc|ccccc}
  \hline
 $N=500$ & only $Y_i$'s centralized & & both $X_i$' and $Y_i$'s centralized &     \\\hline
& Bias& MSE& Bias & MSE \\\hline
$\widehat\beta_{LS}$ & $-1.006888$ & $1.020002$ &$-0.01953344$  &0.01568506 \\
  \hline
\end{tabular}
\end{table}

\begin{figure}
\begin{minipage}{4cm}
\begin{tabular}[b]{c}
{\includegraphics[width=5cm,height=7.2cm]{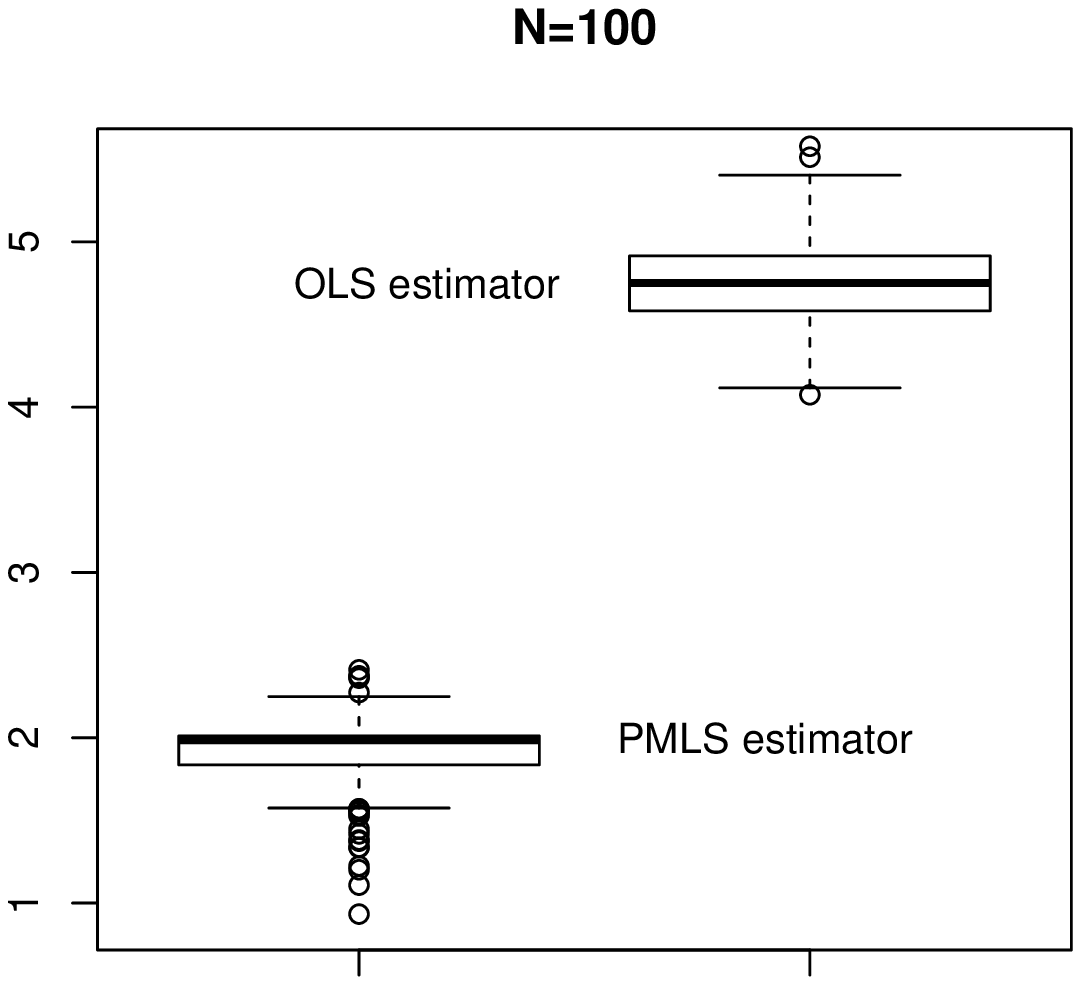}}
\end{tabular}
\end{minipage}
\begin{minipage}{4cm}
\begin{tabular}[b]{c}\ \
{\includegraphics[width=5cm,height=7.5cm]{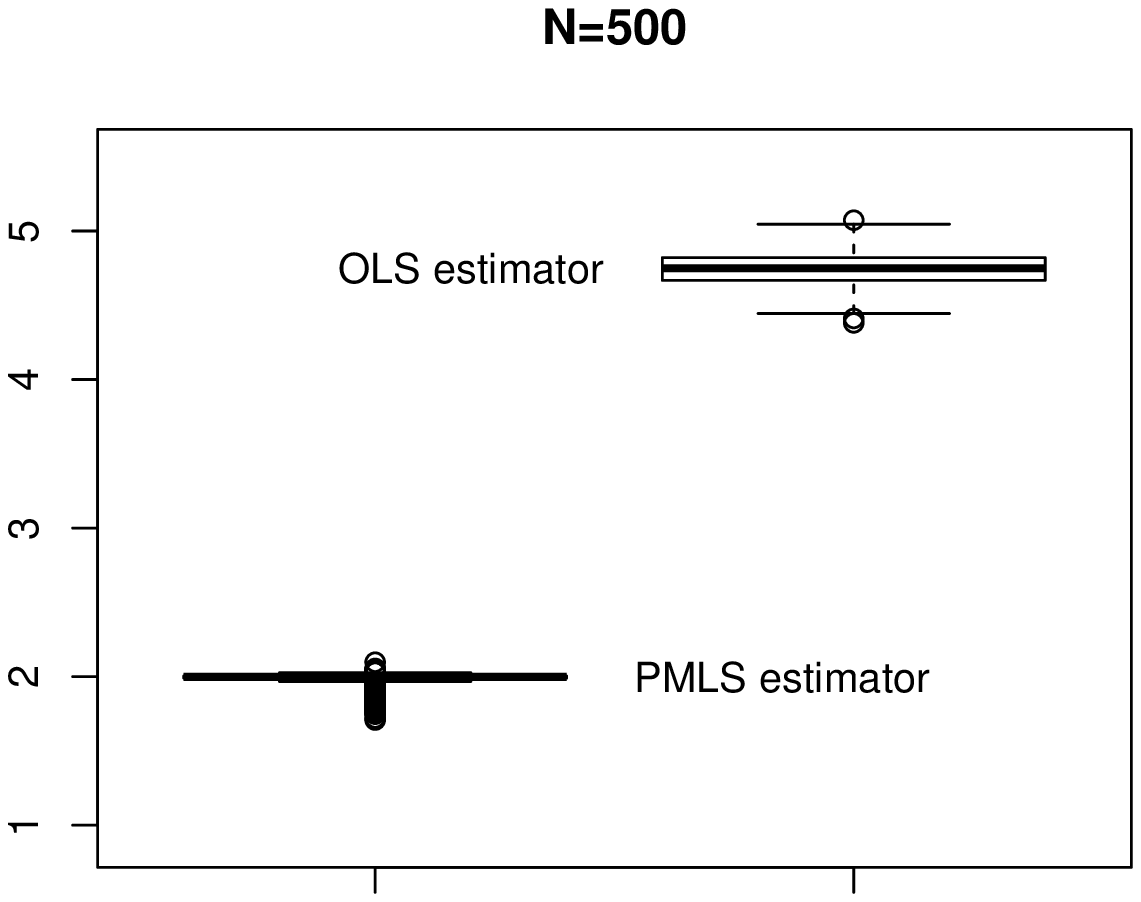}}
\end{tabular}
\end{minipage}
\begin{minipage}{4cm}
\begin{tabular}[b]{c}\ \ \ \ \ \
{\includegraphics[width=4.6cm,height=7cm]{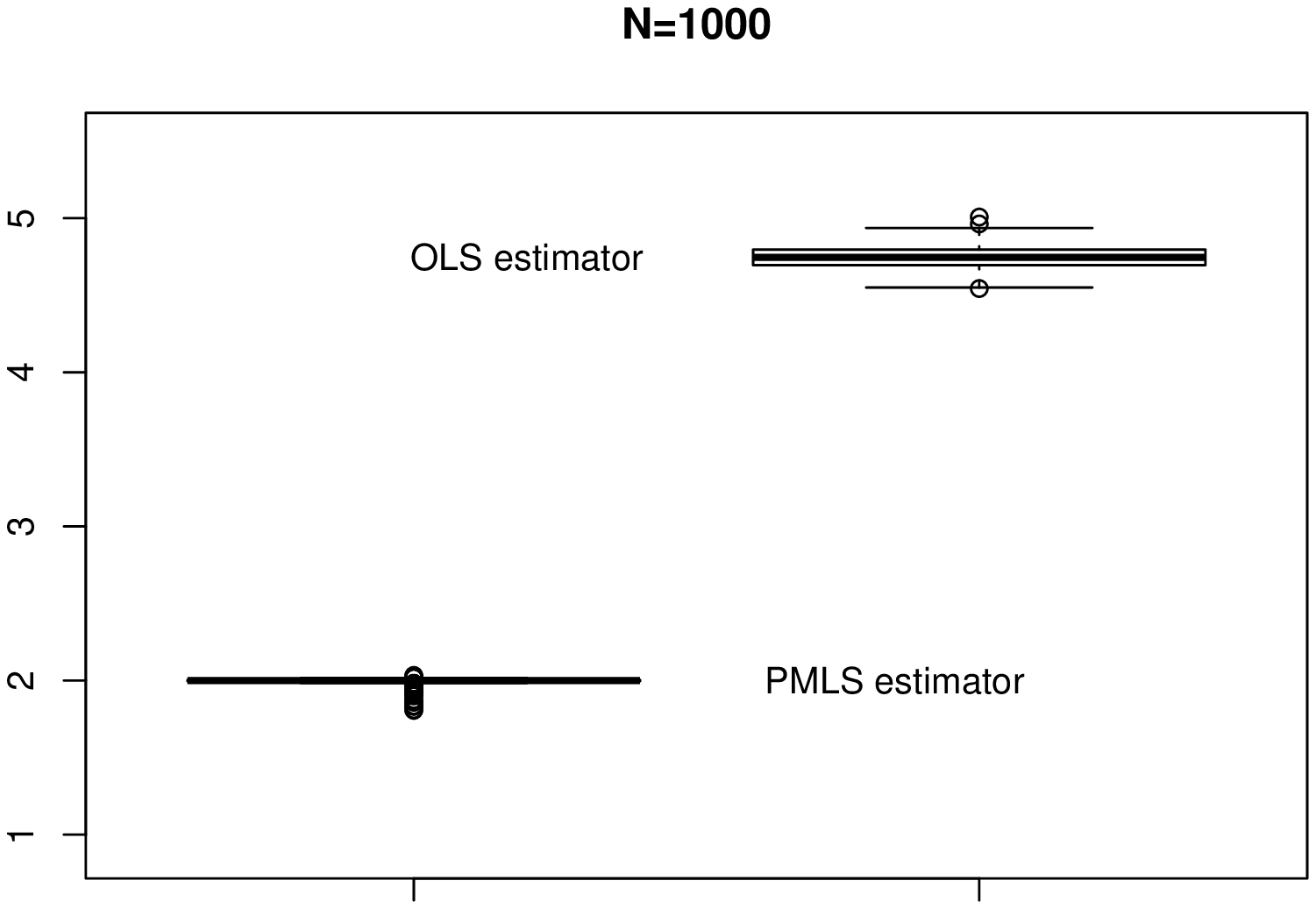}}
\end{tabular}
\end{minipage}\vspace{-6ex}
\caption{The boxplots of the PMLS estimators and the OLS estimators for $\beta$ in Experiment 2 with the true $\beta=2$.} \label{fig:5}
\end{figure}

\begin{figure}
\begin{minipage}{4cm}
\begin{tabular}[b]{c}
{\includegraphics[width=5cm,height=7.2cm]{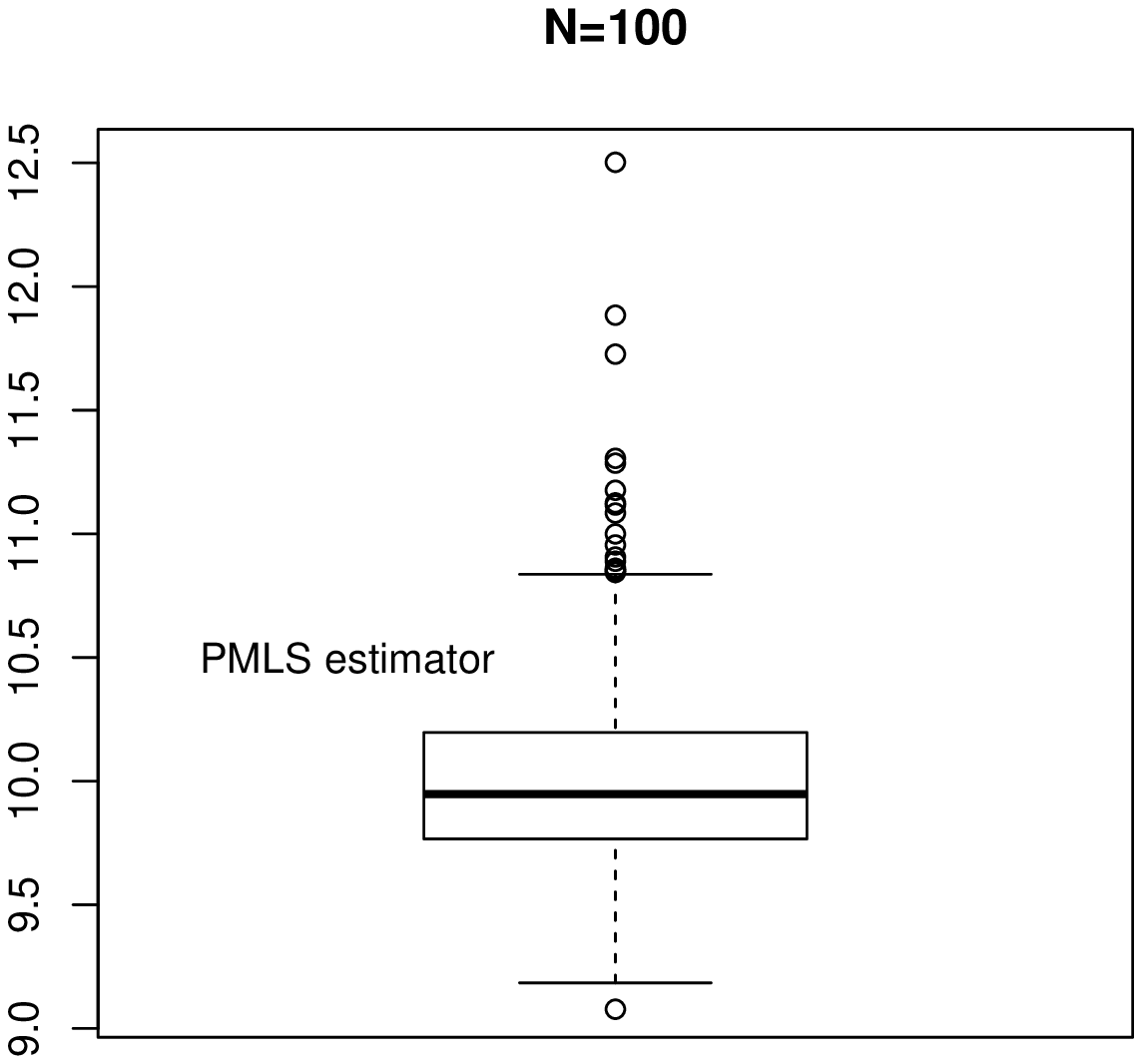}}
\end{tabular}
\end{minipage}
\begin{minipage}{4cm}
\begin{tabular}[b]{c}\ \
{\includegraphics[width=5cm,height=7.2cm]{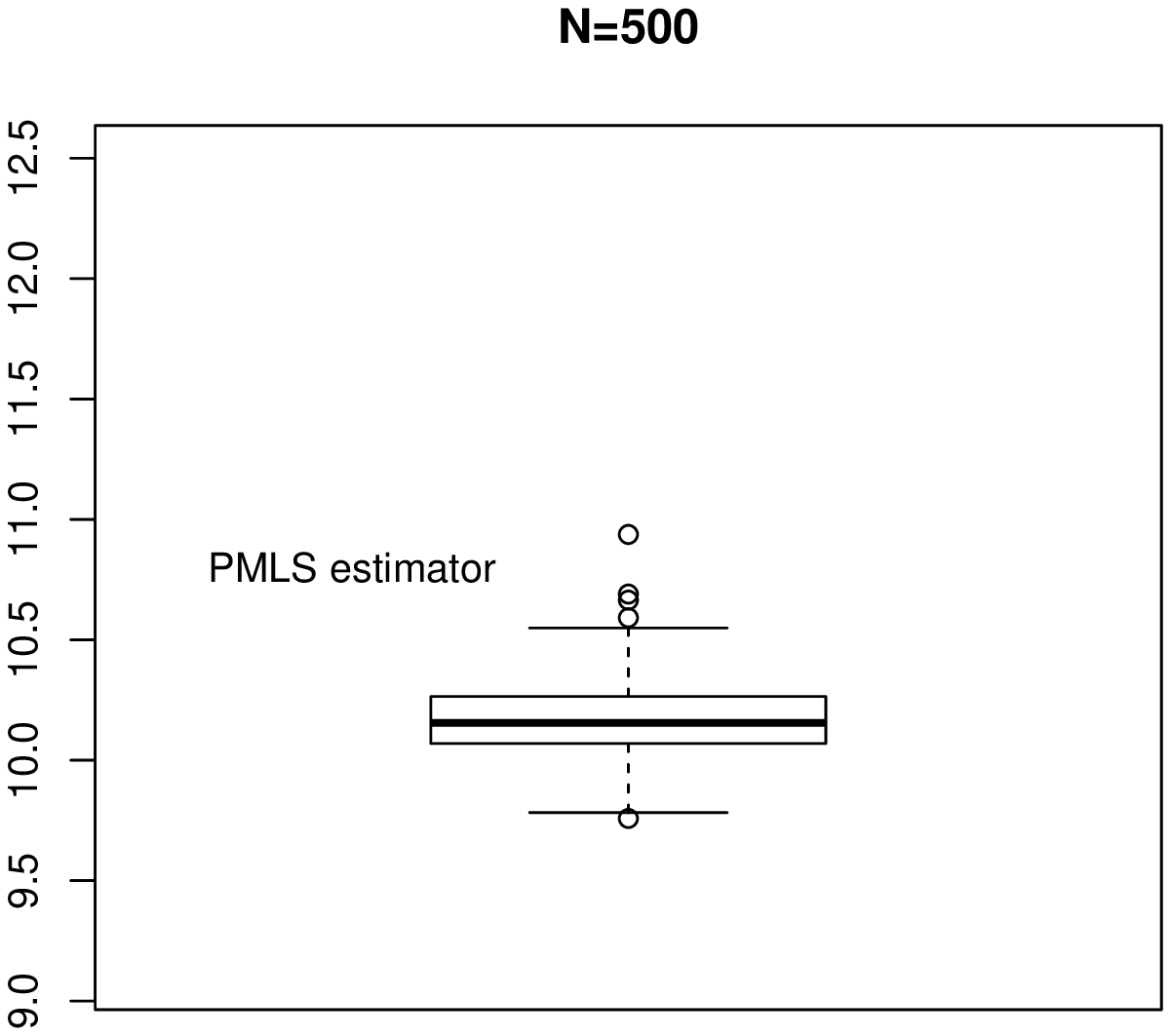}}
\end{tabular}
\end{minipage}
\begin{minipage}{4cm}
\begin{tabular}[b]{c}\ \ \ \ \ \
{\includegraphics[width=5cm,height=6.9cm]{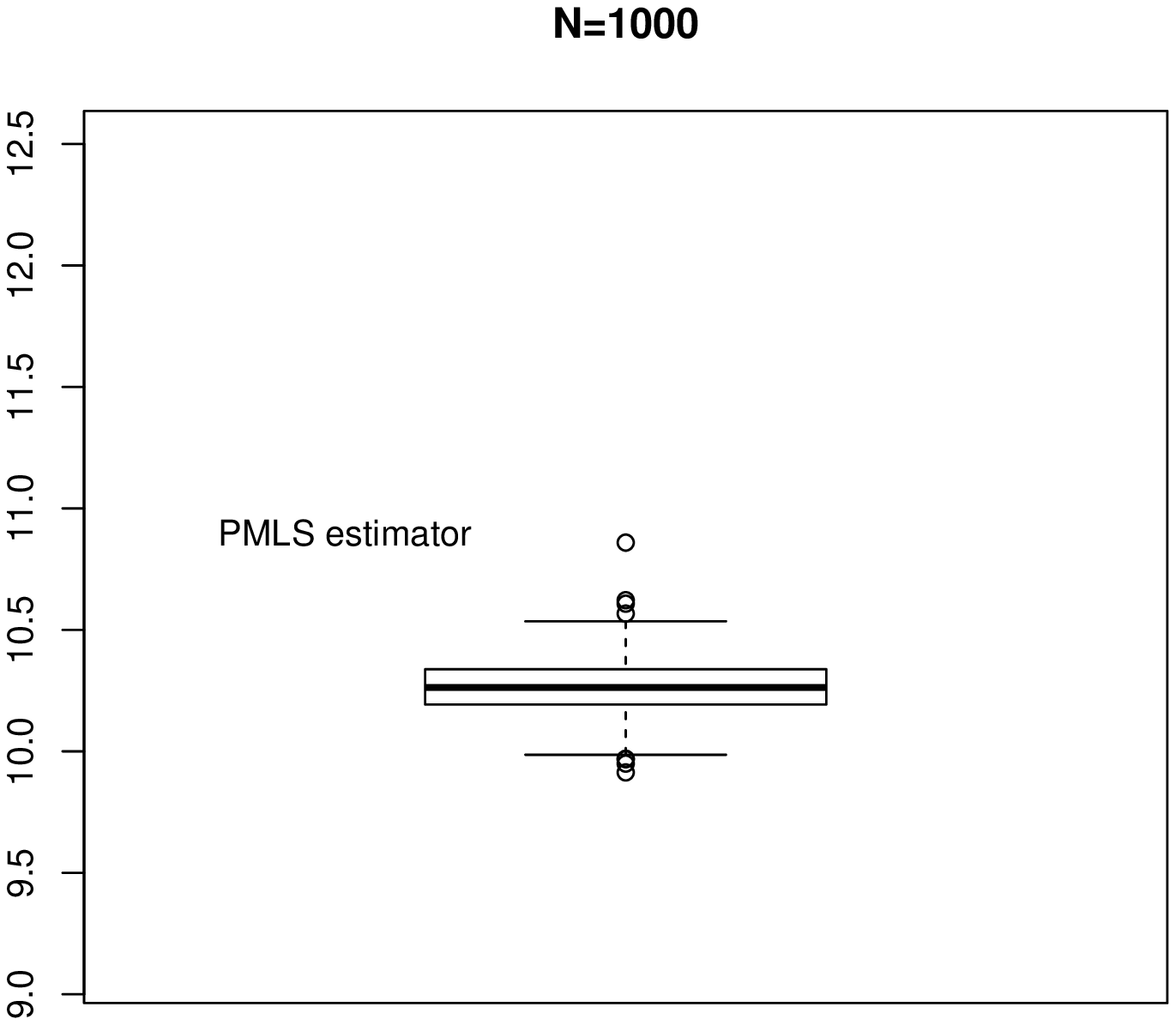}}
\end{tabular}
\end{minipage}\vspace{-6ex}
\caption{The boxplots of the PMLS estimators for $\overline\mu$ in Experiment 2 with the true $\overline\mu=10$.} \label{fig:6}
\end{figure}

\begin{table}
\caption{Estimation bias and MSE in Experiment 3}
\label{tab:5} \vspace{0.3cm} \center
\begin{tabular}{c||cccccccc}
  \hline
 parameters & $N$ & 100& & 500 & &1000& \\
& criterions& Bias& MSE & Bias& MSE & Bias& MSE \\\hline
$\beta_1$ & PMLS & $0.05841$ &0.03646 & $0.01916$& 0.00362&$0.01247$& 0.00169  \\
& OLS &$0.90110$ &0.92334 & $0.91106$& 0.85260& $0.91335$&0.84461\\
  \hline
$\beta_2$ & PMLS & $0.12348$ &0.05200 & $0.03135	$& 0.00495&$0.02113	$& 0.00282  \\
& OLS &$1.85004$ &3.45390 & $1.83649$& 3.38013& $1.83498$&3.37050\\
  \hline
$\overline\mu$   & PMLS & $-0.33971$ &0.34478 & $0.05134$&0.04601&0.18741& 0.06053\\
  \hline
\end{tabular}
\end{table}

\begin{figure}
\begin{minipage}{4cm}
\begin{tabular}[b]{c}
{\includegraphics[width=5cm,height=7.2cm]{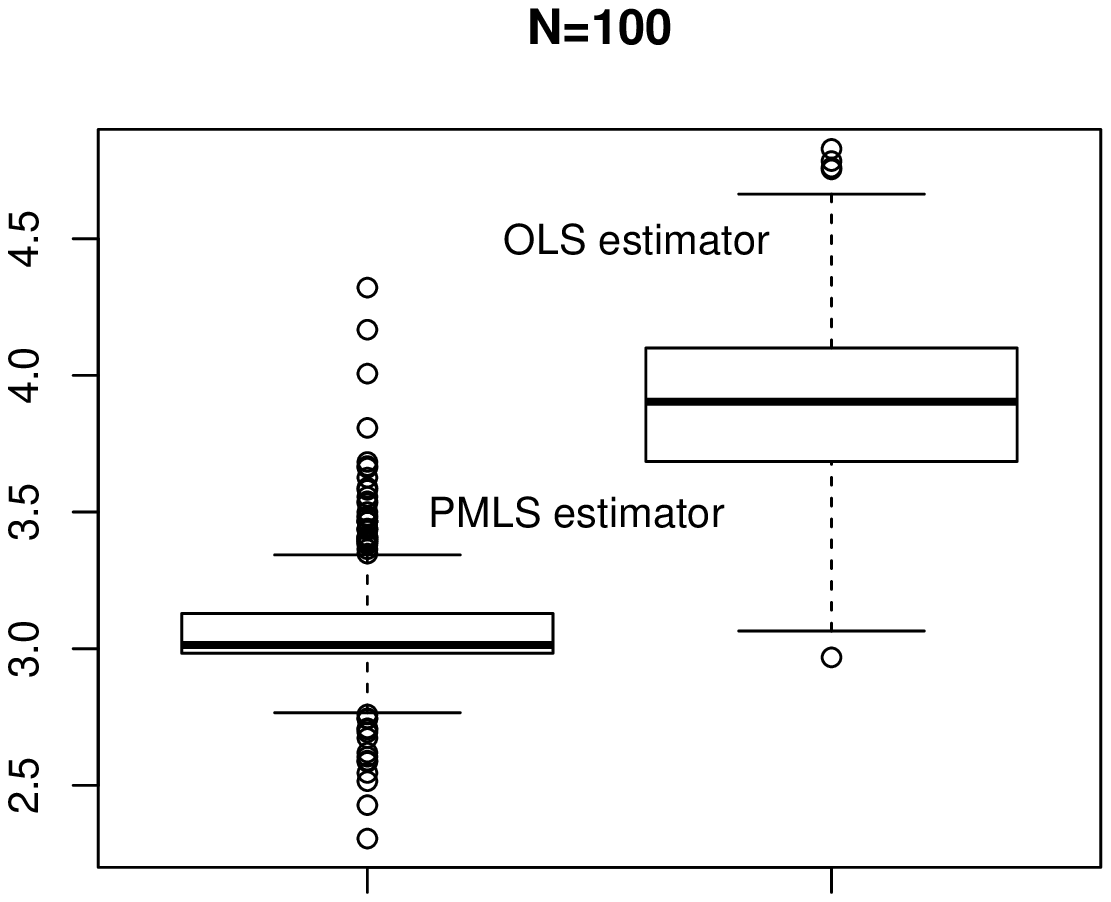}}
\end{tabular}
\end{minipage}
\begin{minipage}{4cm}
\begin{tabular}[b]{c}\ \
{\includegraphics[width=5cm,height=7.2cm]{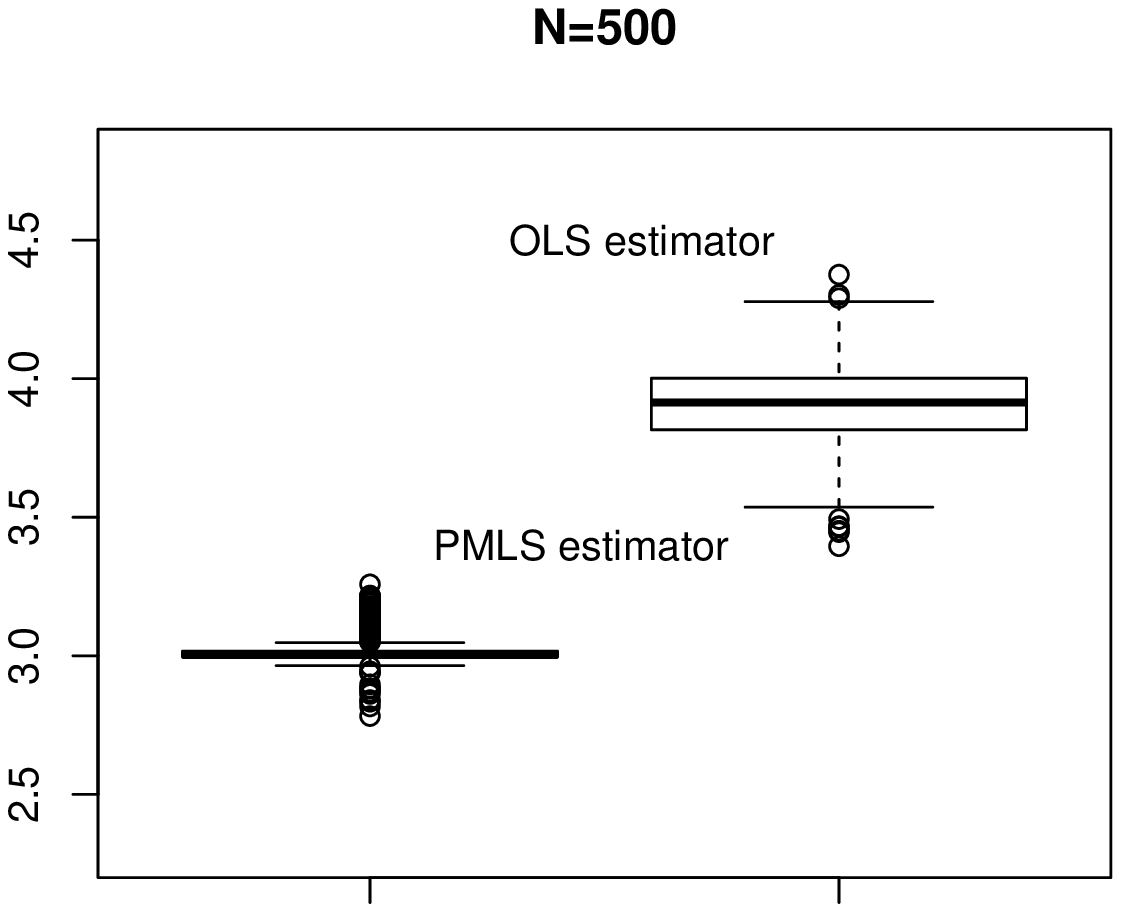}}
\end{tabular}
\end{minipage}
\begin{minipage}{4cm}
\begin{tabular}[b]{c}\ \ \ \
{\includegraphics[width=5cm,height=7.2cm]{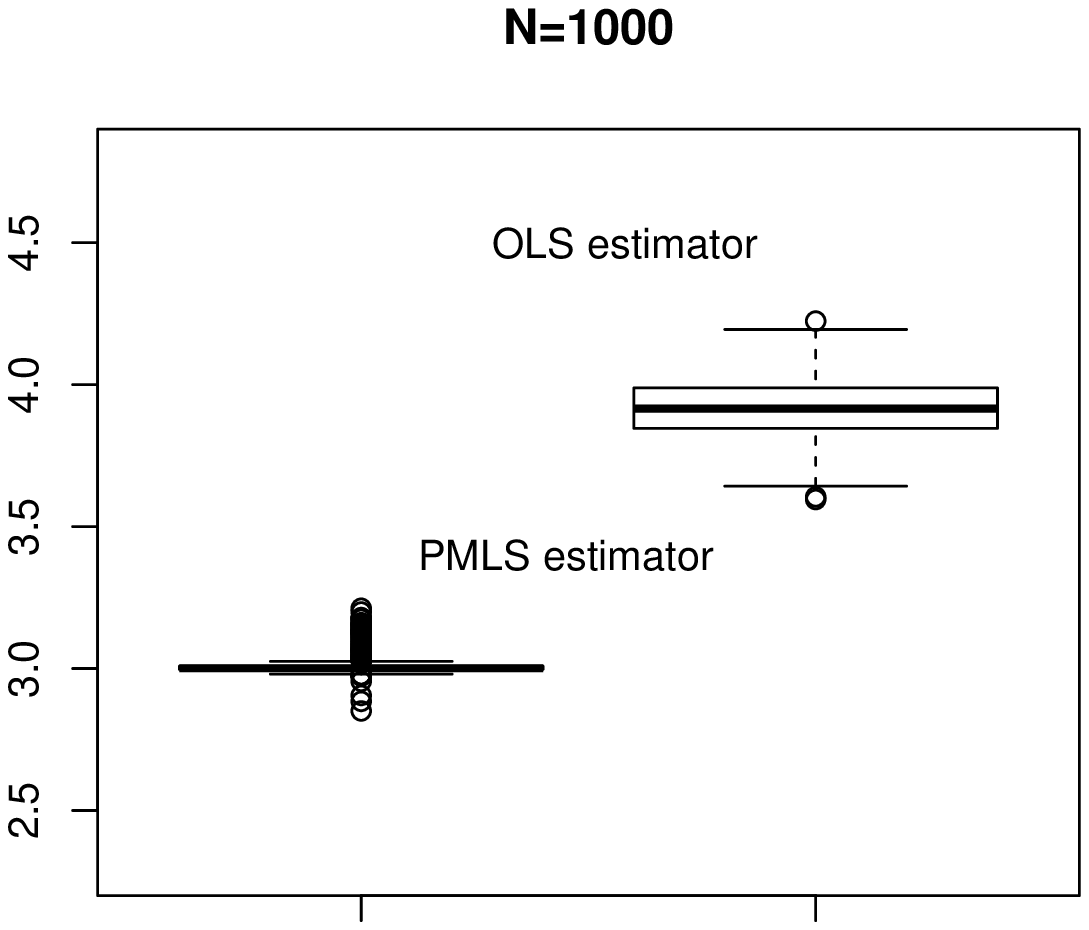}}
\end{tabular}
\end{minipage}\vspace{-6ex}
\caption{The boxplots of the PMLS estimators for $\beta_1$ in Experiment 3 with the true $\beta_1=3$.} \label{fig:7}
\end{figure}

\begin{figure}
\begin{minipage}{4cm}
\begin{tabular}[b]{c}
{\includegraphics[width=5cm,height=7.3cm]{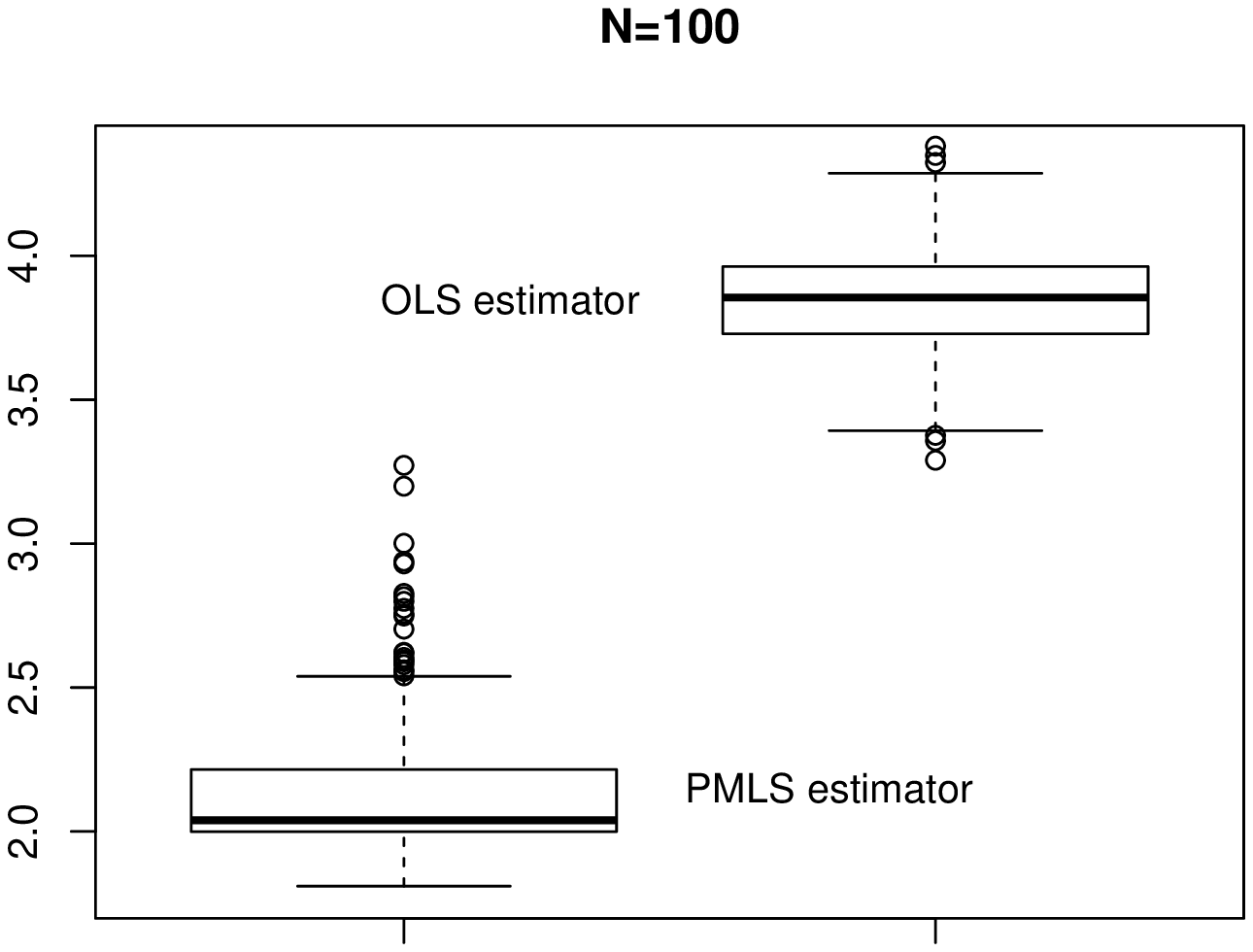}}
\end{tabular}
\end{minipage}
\begin{minipage}{4cm}
\begin{tabular}[b]{c}\ \
{\includegraphics[width=5cm,height=7.3cm]{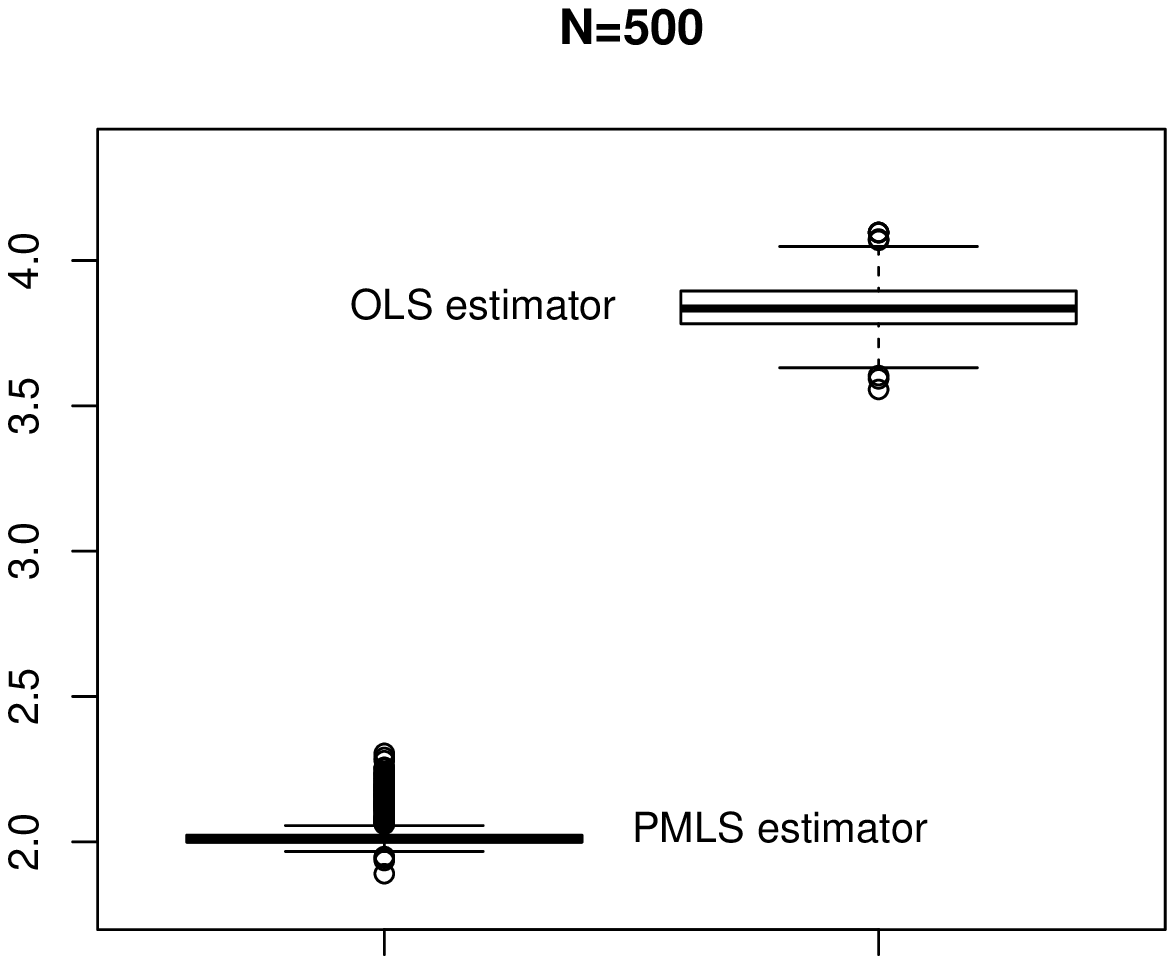}}
\end{tabular}
\end{minipage}
\begin{minipage}{4cm}
\begin{tabular}[b]{c}\ \ \ \ \ \
{\includegraphics[width=5cm,height=7.3cm]{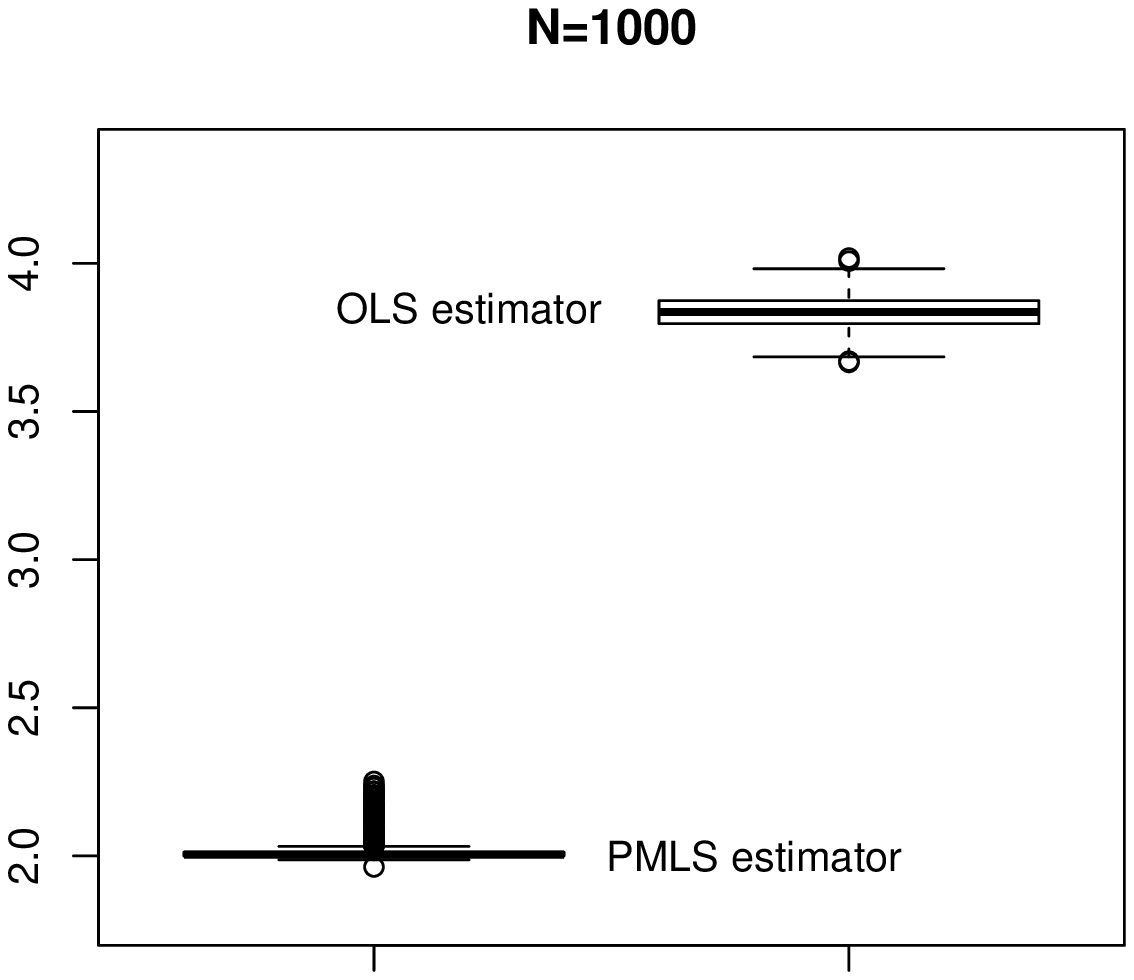}}
\end{tabular}
\end{minipage}\vspace{-6ex}
\caption{The boxplots of the PMLS estimators for $\beta_2$ in Experiment 3 with the true $\beta_2=2$.} \label{fig:8}
\end{figure}

\begin{figure}
\begin{minipage}{4cm}
\begin{tabular}[b]{c}
{\includegraphics[width=5cm,height=7.2cm]{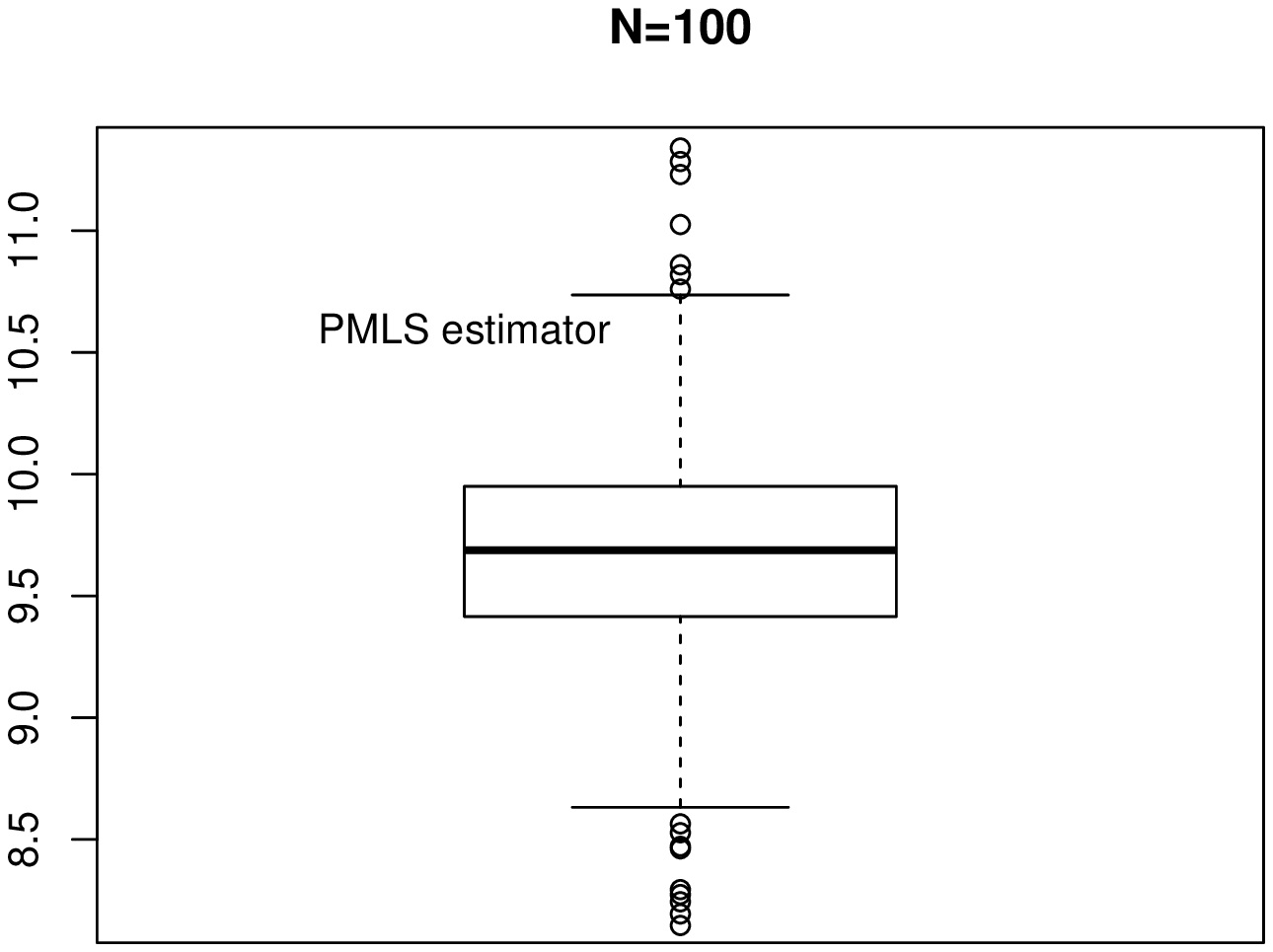}}
\end{tabular}
\end{minipage}
\begin{minipage}{4cm}
\begin{tabular}[b]{c}\ \
{\includegraphics[width=5cm,height=7.3cm]{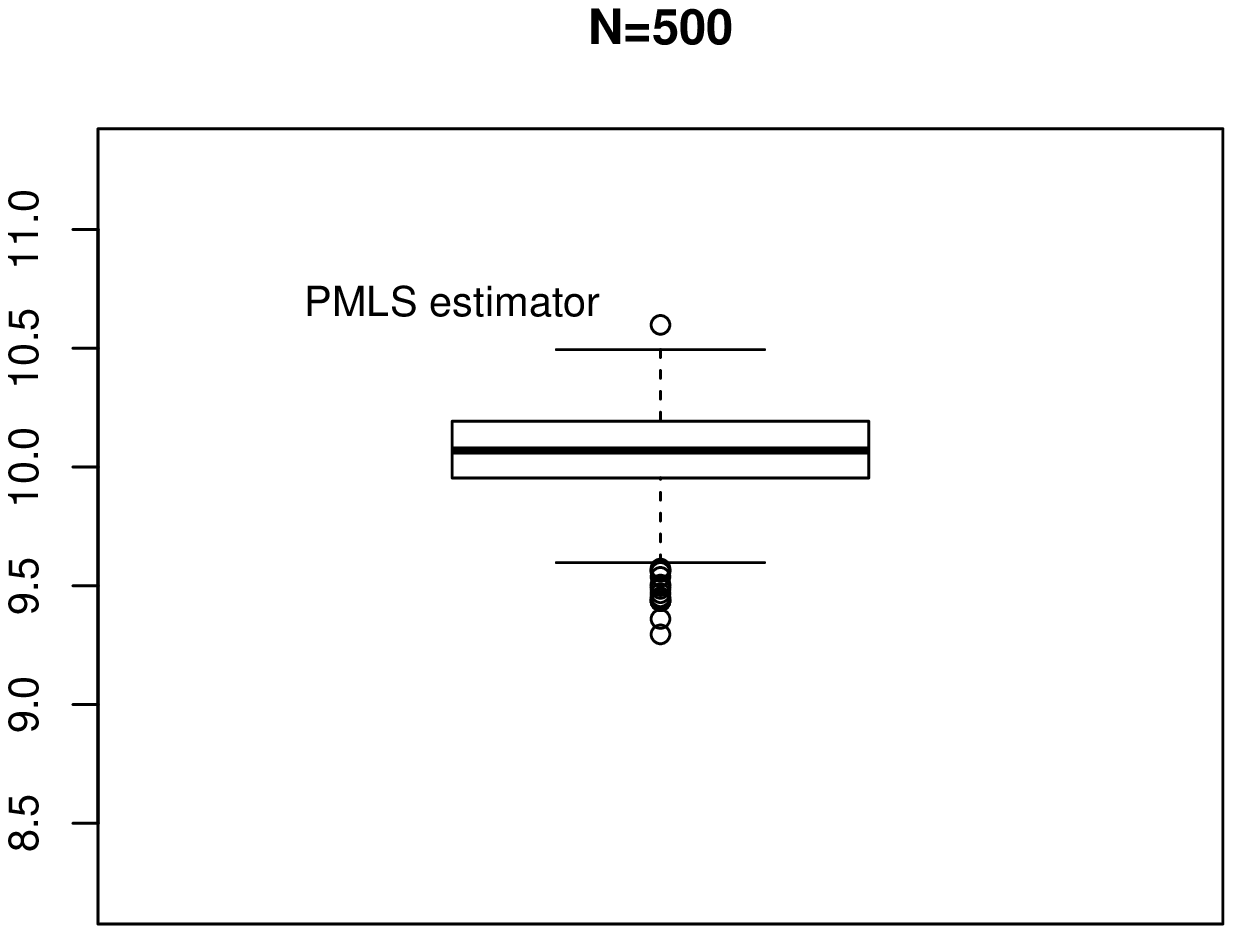}}
\end{tabular}
\end{minipage}
\begin{minipage}{4cm}
\begin{tabular}[b]{c}\ \ \ \ \ \
{\includegraphics[width=5cm,height=7.3cm]{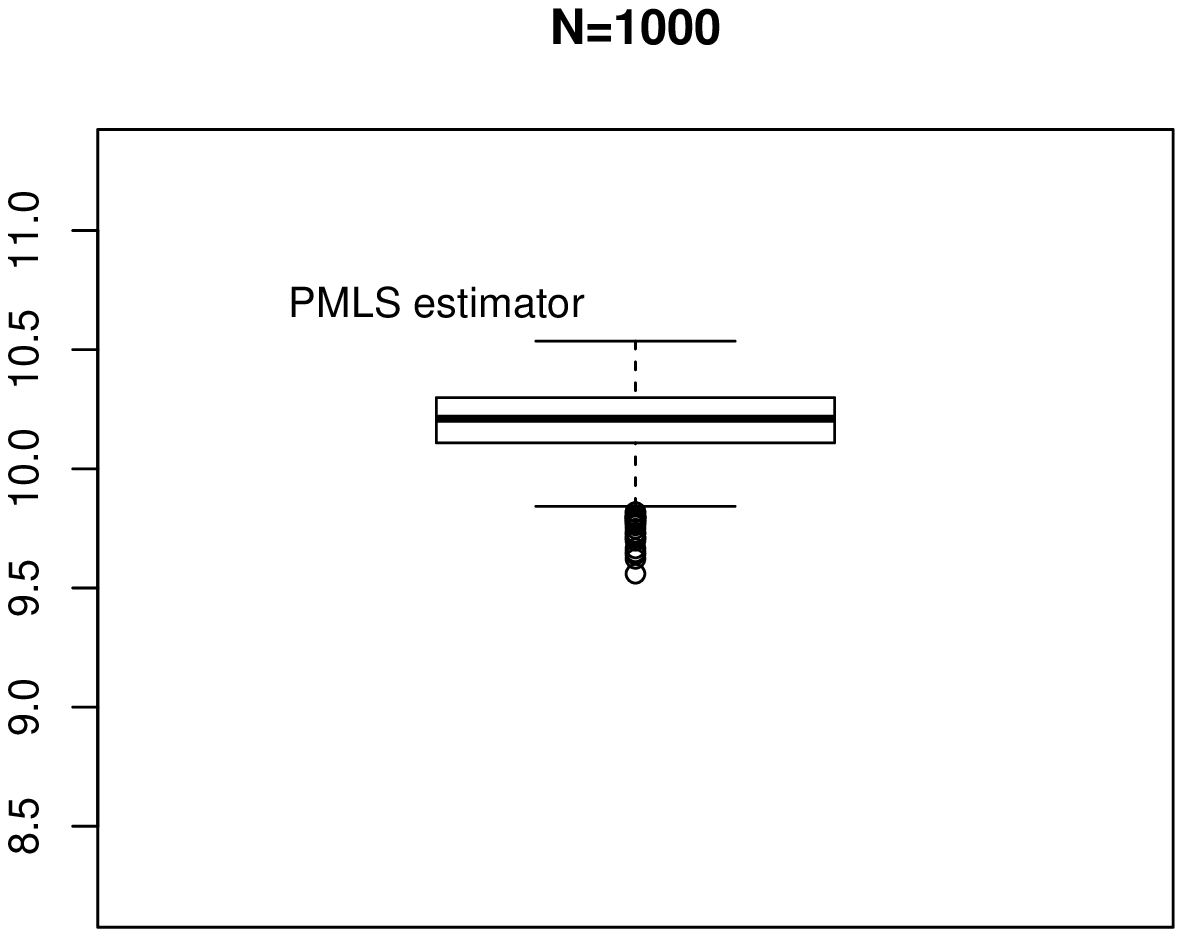}}
\end{tabular}
\end{minipage}\vspace{-6ex}
\caption{The boxplots of the PMLS estimators for $\overline\mu$ in Experiment 3 with the true $\overline\mu=10$.} \label{fig:9}
\end{figure}

\begin{figure}
\begin{minipage}{4cm}
\begin{tabular}[b]{c}
{\includegraphics[width=5cm,height=7.2cm]{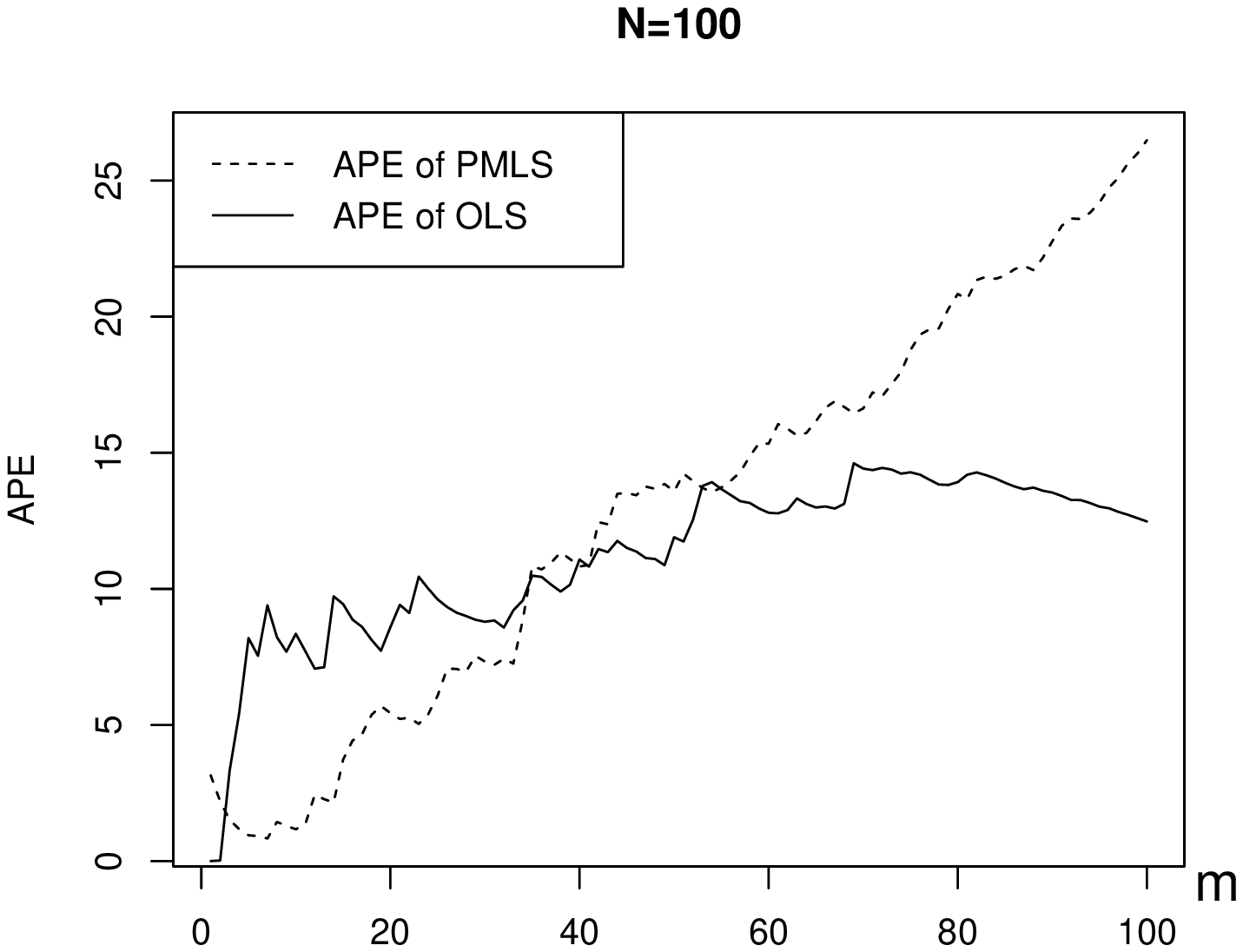}}
\end{tabular}
\end{minipage}
\begin{minipage}{4cm}
\begin{tabular}[b]{c}\ \ \ \
{\includegraphics[width=5cm,height=7.3cm]{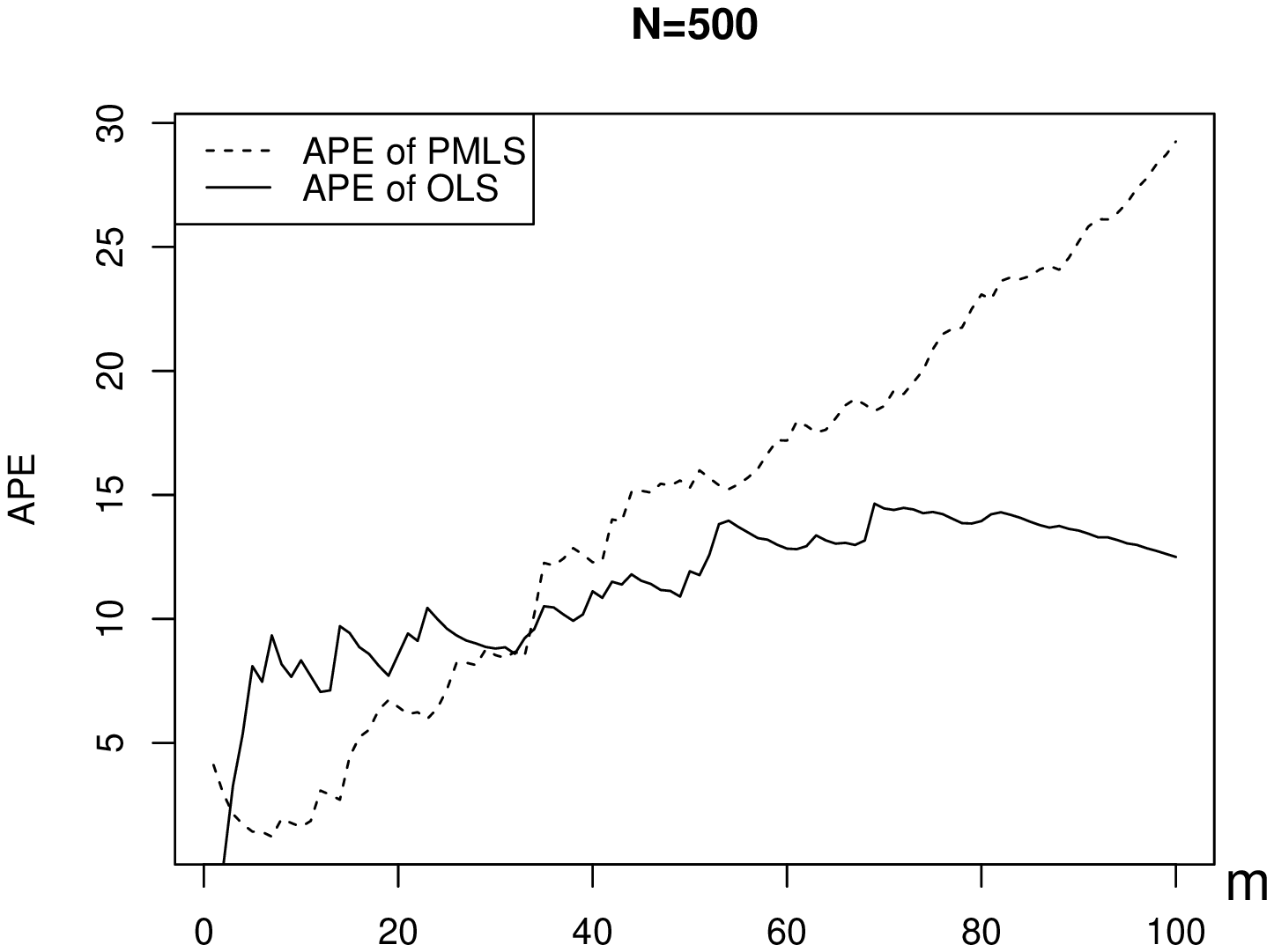}}
\end{tabular}
\end{minipage}
\begin{minipage}{4cm}
\begin{tabular}[b]{c}\ \ \ \ \ \ \ \ \
{\includegraphics[width=5cm,height=7.3cm]{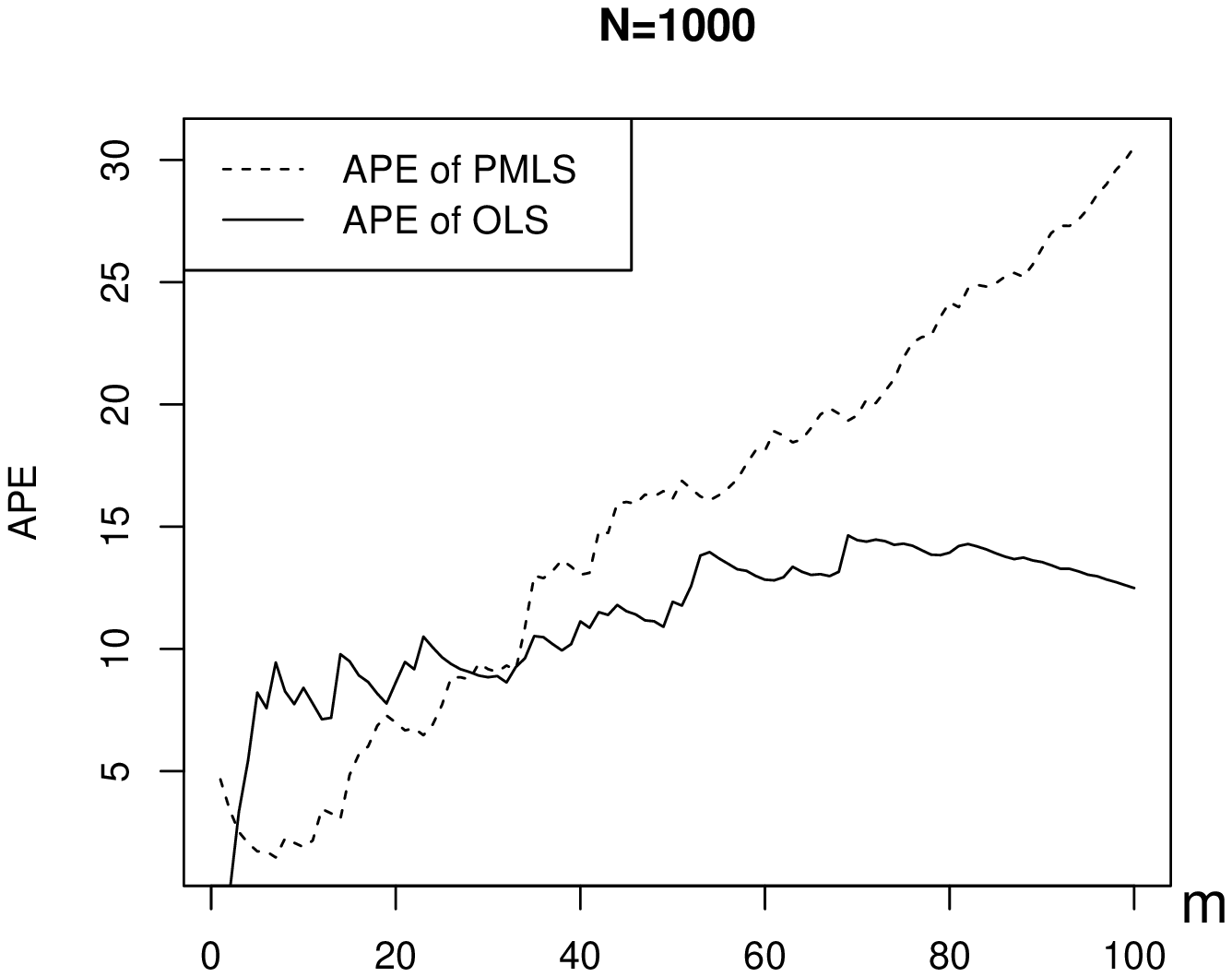}}
\end{tabular}
\end{minipage}\vspace{-6ex}
\caption{The medians of APEs for the first $m$ largest values of $Y$ in Experiment 4.} \label{fig:10}
\end{figure}

\begin{table}
\caption{The APE of predicting all the values of $Y$ in Experiment 4}
\label{tab:6} \vspace{0.3cm} \center
\begin{tabular}{l||cccc}
  \hline
$N$ & 100&  500 & 1000 \\
 \hline
 PMLS &8.498254 & 8.489735 & 8.495854\\
  OLS& 12.47704 &12.49915 & 12.49017  \\
  \hline
\end{tabular}
\end{table}

\begin{figure}
\begin{center}
\begin{tabular}[b]{c}
{\includegraphics[width=9cm,height=7cm]{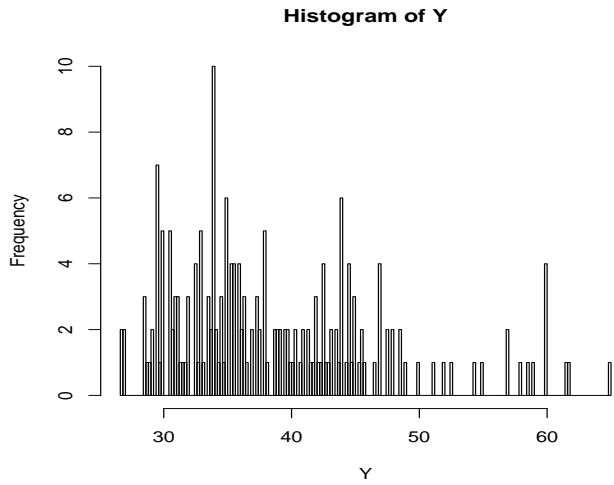}}
\end{tabular}
\vspace{-4ex}\caption{Histogram of Salary.} \label{fig:11}
\end{center}
\end{figure}


\begin{table}
\caption{Parameter estimation and APE for real data}
\label{tab:7} \vspace{0.3cm} \center
\begin{tabular}{l||cccc}
  \hline
parameters &  OLS & PMLS\\
 \hline
$\beta_1$ (Gender) &$-1.314$ &$-3.115$\\
$\beta_2$(PCJob)	&4.532& 5.0824  \\
$\beta_3$(Edu$_1$)&	1.523&	1.969\\
$\beta_4$(Edu$_2$)&	0.086&	0.474\\
$\beta_5$(Edu$_3$)&	$-0.335$&	0.387 \\
$\beta_6$(Edu$_4$)&	$-1.439$&	$-1.692$  \\
$\beta_7$(JobGrd$_1$)&$-36.191$&$-34.143$\\
$\beta_8$(JobGrd$_2$)&$-34.304$&$-31.574$\\
$\beta_9$(JobGrd$_3$)&$-29.392$&$-27.046$\\
$\beta_{10}$(JobGrd$_4$)&$-24.341$&$-22.201$\\
$\beta_{10}$(JobGrd$_5$)&$-17.579$&$-17.210$\\
 $\overline\mu$& --&  68.832\\
 $\underline\mu$& -- &63.628\\
$\beta_0$ (Intercept) & 68.814&--	\\ \hline
APE	&32.615&	32.544\\ \hline
\end{tabular}
\end{table}

\begin{figure}
\begin{center}
\begin{tabular}[b]{c}
{\includegraphics[width=9cm,height=8cm]{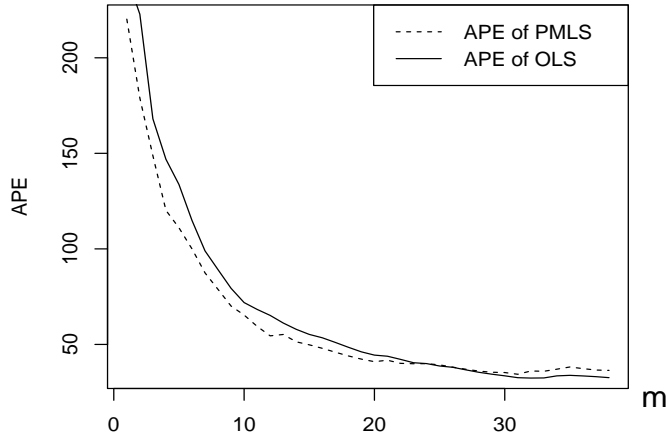}}
\end{tabular}
\vspace{-4ex}\caption{The medians of the APEs for the first $m$ largest values of ``Salary" in real data analysis.} \label{fig:13}
\end{center}
\end{figure}

\begin{figure}
\begin{center}
\begin{tabular}[b]{c}
{\includegraphics[width=9cm,height=8cm]{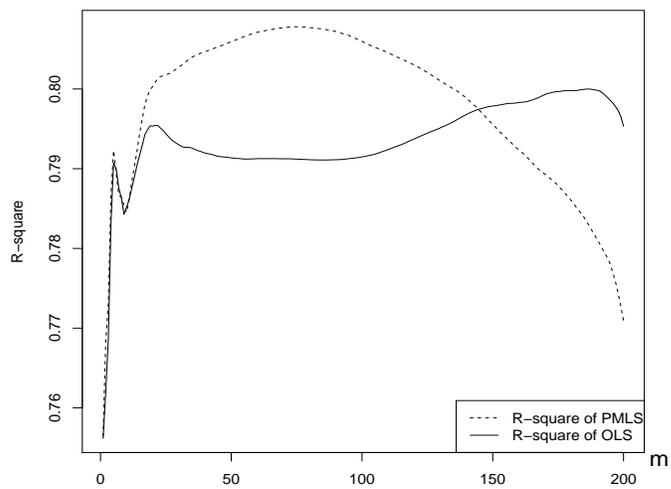}}
\end{tabular}
\vspace{-4ex}\caption{The values of $R^2$ of the first $m$ largest values of ``Salary" in real data analysis.} \label{fig:14}
\end{center}
\end{figure}

\end{document}